\documentclass[preprint,showpacs,preprintnumbers,amsmath,amssymb]{revtex4}
\usepackage{amsmath,amssymb,graphics,epsfig,subfigure}
\usepackage{color}

\begin{document}
\renewcommand{\baselinestretch}{1.3}

\title{Thick branes with a nonminimally coupled bulk-scalar field }

\author{ Heng Guo\footnote{guoh2009@lzu.edu.cn},
         Yu-Xiao Liu\footnote{liuyx@lzu.edu.cn, Corresponding author},
         Zhen-Hua Zhao\footnote{zhaozhh09@lzu.cn},
         Feng-Wei Chen\footnote{chenfw06@lzu.cn}}.\\
        \affiliation{Institute of Theoretical Physics,
              Lanzhou University, Lanzhou 730000,
             People's Republic of China}

\begin{abstract}

In this paper, we investigate thick branes with a
nonminimally coupled background scalar field, whose solution is a
single-kink or a double-kink. The effects of the nonminimal
coupling constant $\xi$ on the structure of the thick branes and the
localization of gravity, fermions, scalars and vectors are
discussed. It is shown that each brane will split into two
sub-branes as increasing the nonminimal coupling constant $\xi$. By
investigating the tensor perturbation equations of gravity and the
general covariant Dirac equation of fermions, we find that both the
gravity zero mode and left-chiral fermion zero mode are localized at
the center of the single-kink branes and localized between the two
sub-branes generated by the double-kink, which indicates that the
constant $\xi$ does not effect the localization of these zero modes.
However, the zero mode of scalars is localized on each sub-brane
(for both single-kink and double-kink branes) when $\xi$ is larger
than its critical value $\xi_0$. The effects of the nonminimal
coupling constant $\xi$ on the resonances of gravity and fermions
with finite lifetime on the branes are also discussed.
\end{abstract}




\maketitle

\section{Introduction}

The idea that our observed four-dimensional Universe might be a
hypersurface, embedded in a higher dimensional space-time (the
bulk), provides new insights into solving the gauge hierarchy and
cosmological constant
problems~\cite{RubakovPLB1983136,VisserPLB1985,ADD,
Randjbar-DaemiPLB1986,rs,Lykken,AntoniadisPLB1990,CosmConst}. In the
Randall-Sundrum (RS) brane-world model, the zero mode of gravity is
localized on the brane, which reproduces the standard Newtonian
gravity on the brane~\cite{rs}. But in this model, the brane is very
ideal because its thickness is neglected. In the most fundamental
theory, there seems to exist a minimum scale of length, thus the
thickness of a brane should be considered in more realistic field
models. For this reason, more natural thick brane scenarios have
been investigated~\cite{De_Wolfe_PRD_2000,PRD_Stojkovic,
Gremm_2000,PRD_koyama,Csaki_NPB_2000,
0005016,PRL845928,CamposPRL2002,Wang_PRD,twobrane,varios,ThickBrane,
Volkas0705.1584,GherghettaPRL2000,Neupane,ThickBraneWeyl,Cvetic,
PRD0709.3552,ThickBrane4,KobayashiPRD2002,RandallJHEP2001,0910.0363,
afonso_plb2006,liu_0911.0269,Brandhuber_JHEP,Diaz_1999,Oda_PRD,
Liu0907.1952,1004.0150,zhong_fRBrane,Singleton,Guo_BentBrane,1009.1684}.
For some comprehensive reviews about thick branes, see Refs.
\cite{0812.1092,0904.1775,0907.3074,1003.1698,brane_book,1004.3962}.

In brane-world theory, the localization of gravity and various bulk
matter fields is a very important issue. In order to recover the
effective four-dimensional gravity, the gravity zero mode should be
localized on branes. In the other hand, various bulk matter fields
should be localized on branes by a natural mechanism, for the
purpose to build up the standard model. Generally, the massless
scalar fields \cite{BajcPLB2000} and graviton \cite{rs} could be
trapped on branes of different types. Spin-1 Abelian vector fields
can be localized on the RS brane in some higher dimensional cases
\cite{OdaPLB2000113} or on the thick de Sitter branes and Weyl thick branes
\cite{Liu0708}. It is important to study the localization problem of
the spin-1/2 fermions. Without introducing the scalar-fermion
coupling, fermions can not be localized on branes in five and six
dimensions \cite{BajcPLB2000,OdaPLB2000113,NonLocalizedFermion,
IchinosePRD2002,Ringeval,
RandjbarPLB2000,KoleyCQG2005,Liu0708,DubovskyPRD2000,0803.1458,
LiuJHEP2007,0901.3543,Liu0907.0910,Koley2009,0812.2638,
LiuJCAP2009,Liu0803,Singleton_fermion,JackiwPRD1976,liu_0909.2312,Guo_jhep,Liu_2010,
zhao_CQG,Fu_KR,zhaoJHEP,FuPRD,PRD_83}. In some cases, there may exist a single bound state
and a continuous gapless spectrum of massive fermion Kaluza-Klein
(KK) modes \cite{ThickBraneWeyl,Liu0708}. In some other cases, one
can obtain finite discrete KK modes (mass gap) and a continuous
gapless spectrum starting at a positive $m^{2}$
\cite{ThickBrane4,LiuJCAP2009,Liu0803}.

Generally, including nonlinear terms of the various curvature
tensors (Riemann, Ricci, Weyl) and nonminimally coupled terms in
the effective action of gravity is a very common trend from quantum
field theory and cosmology. These theories cover $f(R)$ modified
gravity, the Gauss-Bonnet gravity, scalar-tensor gravity, and so on.
On the other hand, in thick brane scenarios, branes are
made of background scalar fields, so it is natural to study gravity
coupled to background scalars. There are a lot of investigations of
thick branes in the case of the minimal coupling, but for the case
of the nonminimal coupling, the investigations are limited.
Recently, brane-world models with a nonminimally coupled bulk-scalar
field, via an interaction term $-\frac{1}{2}\xi R\phi^{2}$ with
$\phi$ a background scalar field and $\xi$ a nonminimal coupling
constant, have been studied. Static solutions of these models have
been examined in Refs.~\cite{PRD73,PLB621,PRD74,PRD77}. The Newton's
law for brane models with a nonminimally coupled bulk scalar field
was investigated in Ref. \cite{PRD76}. The conditions for localization of
gravity for thick brane-worlds with nonminimally coupled term and
Gauss-Bonnet term were discussed in Ref. \cite{1105.5479}. In Ref.
\cite{0811.4253}, the effective dark energy of the brane-Universe
acquires a dynamical nature, as a result of the nonminimal coupling
which provides a mechanism for an indirect ``bulk-brane interaction"
through gravity.

Furthermore, in the brane-world theory, the continuous deformation from
a single brane to two sub-branes, by varying parameters, is called the
phenomenon of brane splitting \cite{CamposPRL2002,twobrane,zhaoJHEP,FuPRD}.
Generally, the single-kink background scalar field
can generate the single-brane, and the double-kink scalar field can also
result in the double-brane. In this paper, we find that increasing
the nonminimal coupling constant $\xi$ can also make the brane splitting
from a single-brane to two sub-branes for the single-kink background scalar field.
The fermion localization on a split brane has been studied in Refs. \cite{zhaoJHEP,FuPRD,PRD_83}.
Here, for the thick branes generated by a nonminimally coupled background scalar,
we find that the different fields are localized on different positions in the extra dimension.
This situation is similar to the so-called ``split fermion" model \cite{PRD61,Dai_PRD},
which offers the solution to the fast proton decay.

In this paper, we are interested in investigating the structure of
thick branes generated by a nonminimally coupled background scalar,
and the effects of the nonminimal coupling constant on the
localization of gravity and various matter fields. Two general cases
of the background scalar field are studied, which are set as a
single-kink and a double-kink. We find that the structure of the
thick branes is very interesting, and it is related to the
nonminimal coupling constant $\xi$. For the single-kink scalar
case, as $\xi$ becomes larger, the single brane will split into two
sub-branes and the distance of the two sub-branes will increase. For
the double-kink scalar case, there are two sub-branes located at
both sides of $z=0$, and as $\xi$ becomes larger, the distance of
the two sub-branes will also increase. Further, the effects of the
nonminimal coupling constant on the localization of gravity and
various matter fields are investigated. Comparing to the case of the
minimal coupling, the localization of the zero modes of gravity and
various matter fields are similar, however, the behavior of the
scalar zero mode is different. With the increase of $\xi$, the
scalar zero mode is localized first on the center of the two
sub-branes for the single-kink case and between them for double-kink
case, and then on each sub-brane. Furthermore, with the increase of
$\xi$, the resonances of gravity will appear, which correspond to
gravitons with a finite lifetime on the branes. This phenomenon does
not appear in the case of the minimal coupling. The resonances of
fermions also exist, and this is similar to the minimal coupling
case. The effects of the nonminimal coupling constant $\xi$ on the
resonances of gravity and fermions are also discussed.

The organization of this paper is as follows: In
Sec.~\ref{SecModel}, the model of thick branes with nonminimally
coupled bulk-scalar field in five-dimensional space-time is
described, and the structure of the branes is also discussed. Then, in
Sec.~\ref{SecGraviton}, we investigate the localization of gravity
on the branes. In Sec.~\ref{SecLocalization}, the localization of
various bulk matter fields is investigated. Finally, the conclusion
is given in Sec.~\ref{SecConclusion}.

\section{The structure of the thick branes with a nonminimally coupled bulk-scalar field }
\label{SecModel}

We start with the following five-dimensional action of thick branes,
which are generated by a real nonminimally coupled scalar field
$\phi$,
\begin{equation}
 S  =  \int d^5 x \sqrt{-g}\left [ F(\phi)
     R-\frac{1}{2}
     g^{MN}\partial_M \phi \partial_N \phi - V(\phi) \right ],
\label{action}
\end{equation}
where $R$ is the five-dimensional scalar curvature and $F(\phi)$ is
chosen as
\begin{eqnarray}
F(\phi)=\frac{1}{2\kappa_{5}^{2}}(1-\kappa_{5}^{2}\xi\phi^{2}),
\end{eqnarray}
$\kappa_5^2=8 \pi G_5$ with $G_5$ the five-dimensional Newton constant
and $\xi$ is a dimensionless coupling constant. The five-dimensional
cosmological constant has been included in the scalar potential
$V(\phi)$. Here, $F(\phi)$ should be positive, and it is clear that
the standard thick brane action is recovered when the coupling
constant $\xi=0$.

The Einstein equations corresponding to the action (\ref{action})
are expressed as follows
\begin{eqnarray}\label{Einstein_Eq}
 R_{MN}-\frac{1}{2}g_{MN} R = T_{MN}
\end{eqnarray}
with $T_{MN}$ the energy-momentum tensor for the scalar field:
\begin{eqnarray}\label{Tmn}
 \!\!\!\!\!\!
 T_{MN}&=& \partial_{M}\phi\partial_{N}\phi-
           g_{MN}\left[\frac{1}{2}g^{PQ}\partial_{P}\phi\partial_{Q}\phi+V(\phi)\right]
           \nonumber\\
       &~&         +2\nabla_{M}\nabla_{N}F(\phi)-2 g_{MN}\Box{F}(\phi)
                 \nonumber\\
       &~& + (1-2F(\phi))\big(R_{MN}-\frac{1}{2}g_{MN} R\big),
\end{eqnarray}
where $\Box$ is the five-dimensional d'Alembertian operator. We have
set $\kappa^2=1$.

The equation of motion for the scalar field reads
\begin{eqnarray}\label{EqScalar}
\Box\phi+ \frac{dF(\phi)}{d\phi} R-\frac{d V(\phi)}{d\phi}=0.
\end{eqnarray}
The above equation and the Einstein equations (\ref{Einstein_Eq})
are not independent \cite{PRD73}.

The line-element for the background space-time describing a thick
flat brane is assumed as
\begin{eqnarray}\label{linee}
 ds^2 =
  g_{MN}dx^{M}dx^{N}
     = \text{e}^{2A(y)}\eta_{\mu\nu}dx^\mu dx^\nu + dy^2,
\end{eqnarray}
where $\text{e}^{2A(y)}$ is the warp factor,
$\eta_{\mu\nu}=\text{diag}(-1,+1,+1,+1)$ is the Minkowski metric,
and $y$ stands for the extra coordinate. We suppose that the scalar
field is considered to be a function of $y$ only, i.e.,
$\phi=\phi(y)$. From Eqs (\ref{Einstein_Eq})-(\ref{linee}), we can
obtain the following equations:
\begin{subequations}\label{EinsteinEq2}
\begin{eqnarray}
\label{EinsteinEq2a}
 3(1-\xi\phi^{2})(A''+2A'^{2})+(\frac{1}{2}-2\xi)\phi'^{2}+V(\phi)
       -2\xi\phi\phi''-6\xi A'\phi\phi'&=&0, ~~~~ \\
\label{EinsteinEq2b}
  6(1-\xi\phi^{2})A'^{2}-\frac{1}{2}\phi'^{2}+V(\phi)-8\xi A'\phi\phi'&=&0, \\
\label{EqScalar2}
 -\phi''-4A'\phi'-\xi(8A''+20A'^{2})\phi+\frac{dV(\phi)}{d\phi}&=&0,
\end{eqnarray}
\end{subequations}
where the prime stands for the derivative with respect to the extra
coordinate. A static analytical solution, for a narrow range of the
coupling constant values $0<\xi<\frac{1}{6}$, has been studied in
Ref. \cite{PRD74}. In this paper, we will investigate the effect of
the coupling constant $\xi$ on the brane structure and the
localization of gravity and various spin fields in some general
situations.

For the sake of convenience of obtaining the mass-independent
localization potential for gravitons, we will follow Ref. \cite{rs}
and change the metric given in (\ref{linee}) to the conformally flat
form
\begin{eqnarray}\label{line_z}
 ds^{2}= \text{e}^{2A(z)}(\eta_{\mu\nu}dx^\mu dx^\nu + dz^2)
\end{eqnarray}
by performing the coordinate transformation
\begin{equation}\label{transformation}
dz=\text{e}^{-A(y)}dy.
\end{equation}
%
The equations of the motion
for the background scalar field $\phi(z)$ and the warp factor $A(z)$
in the $z$ coordinate can be written as
\begin{subequations}\label{Einstein_Eq_z}
\begin{eqnarray}
\label{Einstein_Eq_z1}
 3(1-\xi\phi^{2})(A'^{2}+A'')+\frac{1}{2}\phi'^{2}+\text{e}^{2A}V(\phi)
     -2\xi\phi'^{2} -2\xi\phi\phi''- 4\xi\phi\phi'' &=&0, ~~~~~~\\
\label{Einstein_Eq_z2}
 6(1-\xi\phi^{2})A'^{2}-\frac{1}{2}\phi'^{2}+\text{e}^{2A}V(\phi)
   -8\xi\phi A'\phi' &=&0,  \\
\label{Einstein_Eq_z3}
 \phi''+3A'\phi'+4\xi\phi(3A'^{2}+2A'')-\text{e}^{2A}\frac{dV(\phi)}{d\phi}
   &=& 0.
\end{eqnarray}
\end{subequations}
As we have already mentioned, the above equations are not independent.
Hence, we choose to solve Eq. (\ref{Einstein_Eq_z2}) and the
following equation, which can be obtained from (\ref{Einstein_Eq_z1}) and (\ref{Einstein_Eq_z2}),
\begin{eqnarray}\label{EinsteinEqz}
 3(1-\xi\phi^{2})(A''-A'^{2})+4\xi\phi A'\phi' \nonumber\\
 + (1-2\xi)\phi'^{2}
   -2\xi\phi\phi''=0.~~
\end{eqnarray}
By considering that the energy density $T_{00}(z)$ should vanish as $z\rightarrow\pm\infty$, the background scalar field
should satisfy $\phi(z\rightarrow\pm\infty)<\infty$ and
$\phi'(z\rightarrow\pm\infty)<\infty$, so the background scalar
field is naturally considered as a kink solution
\begin{eqnarray}\label{BackgroundScalar}
\phi(z)= \phi_{0}\tanh^{k}(b z),
\end{eqnarray}
where $\phi_{0}$ and $b$ are positive real parameters, and the
parameter $k$ is an positive odd number. Here we just only consider
two cases: the $k=1$ single-kink solution and the $k=3$ double-kink
solution. The background scalar has the following behavior at $z=0$
and $z\rightarrow\pm\infty$:
\begin{eqnarray}
\phi(0)=0, \quad\quad \phi(z\rightarrow\pm\infty)=\pm\phi_{0}.
\end{eqnarray}
Because $F(\phi)>0$ and $\phi\leq\phi_{0}$, the coupling constant
$\xi<\frac{1}{\phi_{0}}$. It is very difficult to solve Eq.
(\ref{EinsteinEqz}) by analytical method, so we will numerically
solve it. First, we will analyze the asymptotic behavior of the warp
factor at $z\rightarrow\pm \infty$. When $z\rightarrow\pm\infty$, we
have for the scalar field that $\phi\rightarrow\pm\phi_{0}$,
$\phi'\rightarrow 0$ and $\phi''\rightarrow 0$. Hence, Eq.
(\ref{EinsteinEqz}) reduces to $A''(z)-A'^{2}(z)=0$ as
$z\rightarrow\pm\infty$, from which the asymptotical form of the
warp factor can be given as
$\text{e}^{A(z)}=\frac{1}{c+b|z|}\simeq\frac{1}{b|z|}$, which is the
same as the RS brane-world. Next, we want to obtain the approximate
solution of Eq. (\ref{EinsteinEqz}) in the vicinity of $z=0$. We
expand the background scalar field $\phi(z)$ at $z=0$:
\begin{eqnarray}
 \phi(z)
  &=&\phi_{0}(b z)^{k}-\frac{1}{3}\phi_{0}k(b z)^{k+2}+\mathcal{O}\big((b z)^{k+4}\big).
\end{eqnarray}

For $k=1$, by using
 $\phi(z)=\phi_{0}(bz)-\frac{1}{3}\phi_{0}(b z)^{3}+\mathcal{O}\big((b z)^{5}\big)$,
the behavior of the warp factor in the vicinity of $z=0$, from Eq.
(\ref{EinsteinEqz}), can be expressed as
\begin{eqnarray}
 A(z)=\frac{1}{6}\phi_{0}^{2}(2\xi-1)(bz)^{2} + \mathcal{O}\big((b z)^{3}\big).
  \label{Azk1}
\end{eqnarray}
We find that the behavior of the warp factor in the vicinity of
$z=0$ is related closely to a critical coupling constant
$\xi_{0}=\frac{1}{2}$. When small $\xi$ with $\xi<\xi_{0}$, the
maximum of the warp factor $\text{e}^{2A(z)}$ is at $z=0$, and this
will correspond to one brane located at $z=0$. When $\xi$ is closer
to $\xi_0$, we will see that the thick brane will split into two
sub-branes. When $\xi>\xi_{0}$, the maxima of the warp factor are at
both sides of $z=0$ and this will result in the increase of the
distance of the two sub-branes.

For $k=3$, the scalar $\phi(z)$ can be expressed as
 $\phi(z)=\phi_{0}(b z)^{3}+\mathcal{O}\big((b z)^{5}\big)$
at $z=0$. And the warp factor can be described as
\begin{eqnarray}
 A(z)=\frac{1}{30}\phi_{0}^{2}(10\xi-3)(bz)^{6}+\mathcal{O}\big((b z)^{7}\big).
   \label{Azk3}
\end{eqnarray}
Here the critical coupling constant is $\xi_{0}=\frac{3}{10}$. The
behavior of the warp factor is the same as the case $k=1$. However,
there are always two sub-branes for any $\xi$ because the scalar is
a double-kink.

\begin{figure*}[htb]
\begin{center}
\subfigure[$k=1$]{\label{fig_WarpFactorI}
\includegraphics[width=7cm]{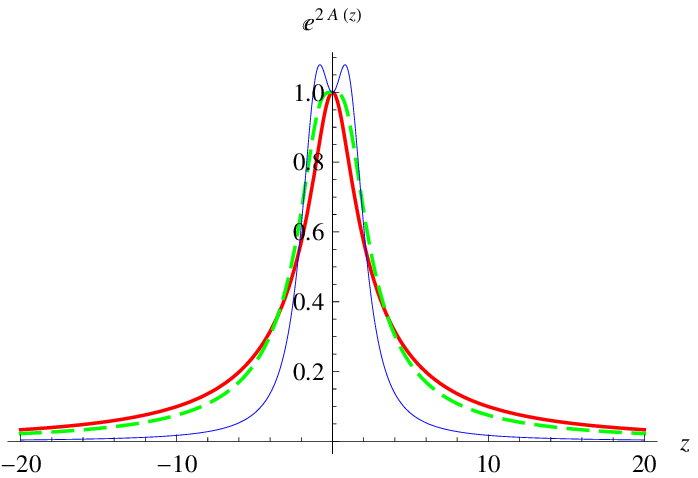}}
\subfigure[$k=1$]{\label{fig_T00I}
\includegraphics[width=7cm]{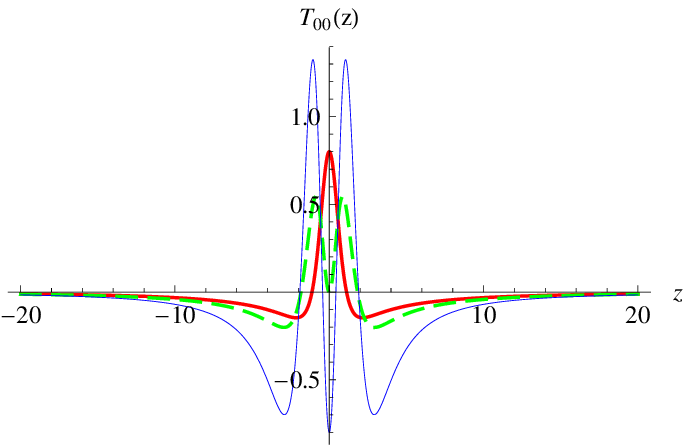}}
\subfigure[$k=1$]{\label{fig_phi1}
\includegraphics[width=7cm]{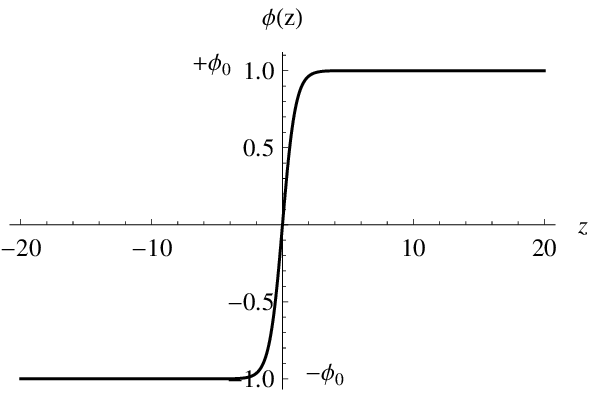}}
\subfigure[$k=1$]{\label{fig_Vphi1}
\includegraphics[width=7cm]{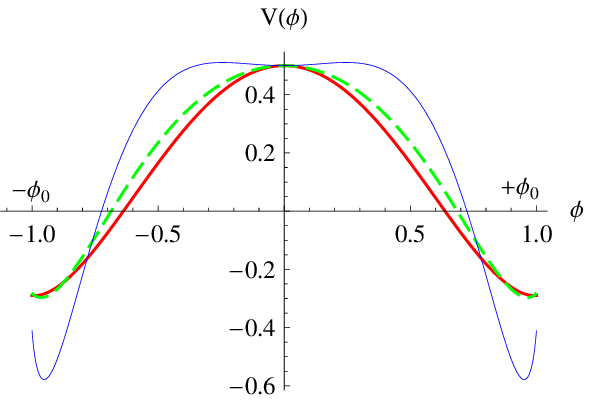}}
\end{center}\vskip -5mm
\caption{The shapes of the warp factor $\text{e}^{2A}$, the energy
density $T_{00}(z)$, the background scalar field $\phi(z)$ and
scalar potential $V(\phi)$. The parameters are set to $\xi=0.1$ for
the thick red line, $\xi=0.5$ for the dashed green line, and
$\xi=0.9$ for the thin blue line. The other parameters are set as
$\phi_{0}=1$, $b=1$, and $k=1$. }
 \label{fig_WarpT00I}
\end{figure*}

\begin{figure*}[htb]
\begin{center}
\subfigure[$k=3$]{\label{fig_WarpFactor2}
\includegraphics[width=7cm]{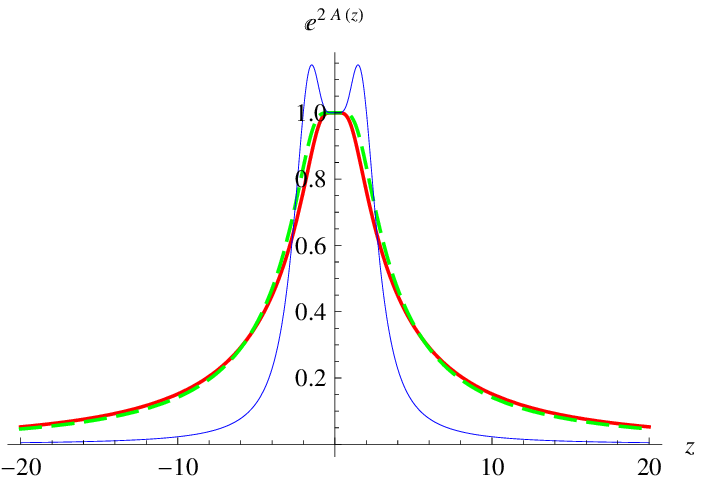}}
\subfigure[$k=3$]{\label{fig_T002}
\includegraphics[width=7cm]{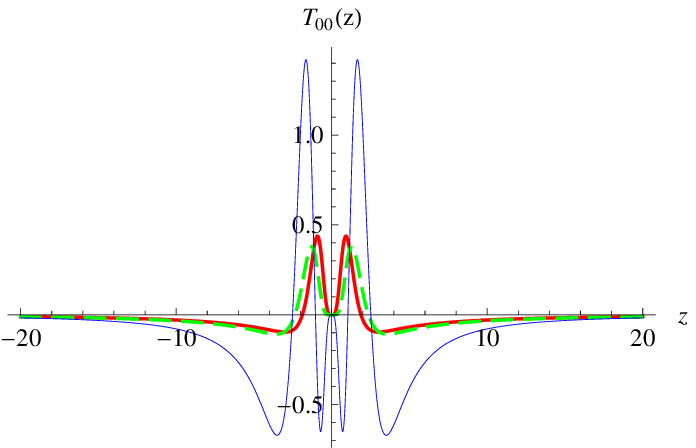}}
\subfigure[$k=3$]{\label{fig_phi2}
\includegraphics[width=7cm]{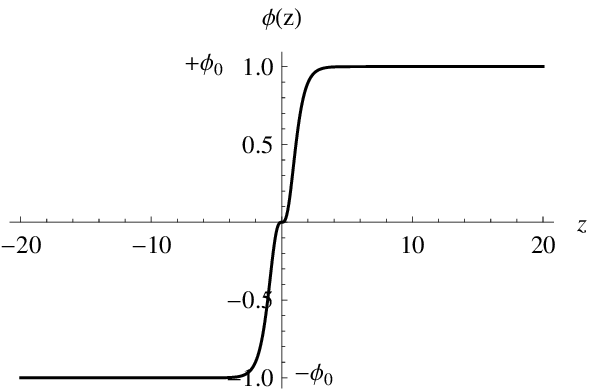}}
\subfigure[$k=3$]{\label{fig_Vphi2}
\includegraphics[width=7cm]{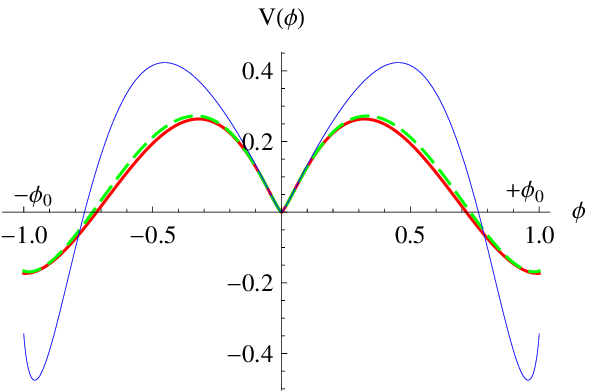}}
\end{center}\vskip -5mm
\caption{The shapes of the warp factor $\text{e}^{2A}$, the energy
density $T_{00}(z)$, the background scalar field $\phi(z)$ and
scalar potential $V(\phi)$. The parameters are set to $\xi=0.1$ for
the thick red line, $\xi=0.3$ for the dashed green line, and
$\xi=0.9$ for the thin blue line. The other parameters are set as
$\phi_{0}=1$, $b=1$, and $k=3$. }
 \label{fig_WarpT00II}
\end{figure*}

Eq. (\ref{EinsteinEqz}) can be solved numerically with the following
initial conditions:
\begin{eqnarray}
 A(0)=A'(0)=0.
\end{eqnarray}
The shapes of the warp factor $\text{e}^{2A}$ and the energy density
$T_{00}(z)$ are shown in Fig.~\ref{fig_WarpT00I} and
Fig.~\ref{fig_WarpT00II} for $k=1$ and $k=3$, respectively. For
$k=1$, when $\xi$ is very small, there is only one brane located at
$z=0$, and when increasing $\xi$, the brane gradually splits two
sub-branes located at both sides of $z=0$. The distance of the two
sub-branes increases with $\xi$. For $k=3$, there are two sub-branes
located at both sides of $z=0$ even for small $\xi$, and the
distance of the two sub-branes also increases with $\xi$.

The scalar potential $V(\phi)$ can also be solved
numerically by Eqs. (\ref{Einstein_Eq_z2}) and (\ref{EinsteinEqz}),
and it is shown in Fig.~\ref{fig_Vphi1} and Fig.~\ref{fig_Vphi2} for
single-kink ($k=1$) and double-kink ($k=3$), respectively. Because
the range of the background scalar $\phi(z)$ is from $-\phi_0$ to
$+\phi_0$, the scalar potential $V(\phi)$ only can be given between
$V(-\phi_0)$ and $V(+\phi_0)$ by numerical method. It is shown that
the scalar potential is a double well potential. When the coupling
constant $\xi=0$, the vacua of the potential are at $\pm\phi_{0}$,
and the background scalar (single- or double-kink) connects two
vacua of the double well potential. When $\xi>0$, the two vacua are
not at $\pm\phi_0$, and this can also be obtained by analyzing the
field equations. From Eq. (\ref{Einstein_Eq_z3}), the following
expression is obtained easily
\begin{eqnarray}
 \label{dvdphi}
 \frac{d V(\phi)}{\phi}=
   \text{e}^{-2A}\big[\phi''+3A'\phi'+4\xi\phi(3A'^{2}+2A'')\big].
\end{eqnarray}
Considering that the background scalar field and the warp factor
have the following asymptotic behavior $\phi(\pm\infty)\rightarrow
\pm\phi_0$, $\phi'(\pm\infty)\rightarrow\pm 4k\phi_0 \;\text{e}^{-2z}$,
$\phi''(\pm\infty)\rightarrow\mp 8k\phi_0 \;\text{e}^{-2z}$, $A(\pm\infty)\rightarrow
-\ln b|z|$, $A'(\pm\infty)\rightarrow -\frac{1}{|z|}$ and
$A''(\pm\infty)\rightarrow \frac{1}{z^{2}}$, the above expression
can be reduced at infinity to
\begin{eqnarray}
 \frac{dV(\phi)}{d\phi}
   \rightarrow \pm 20\xi b^{2}\phi_0,
\end{eqnarray}
and it is clear that the two vacua of the potential $V(\phi)$ are at
$\pm\phi_{0}$ only for $\xi=0$. At $\phi=0$, the scalar potential
$V(\phi)$ has different behaviors for $k=1$ and $k=3$. For $k=1$,
when the coupling constant $\xi$ is small, the scalar potential
$V(\phi)$ has a local maximum at $\phi=0$, which is similar to the
$\phi^{4}$ model in particle physics, and when $\xi$ becomes large
and closes to $\frac{1}{\phi_0}=1$, the potential at $\phi=0$ turns
to a local minimum, which is similar to $\phi^{6}$ model in particle
physics. For $k=3$, $V(\phi)$ always has a local minimum at
$\phi=0$, which is the same as $\phi^{6}$ model.

\section{Localization of gravity on the thick branes}
\label{SecGraviton}

In this section we will investigate the localization of the gravity
on the branes, by the linearized equations for the metric
fluctuations. From the Einstein equations (\ref{Einstein_Eq}), we
can obtain the alternative form of the Einstein equations:
\begin{eqnarray}\label{EinsteinEqform2}
 2F(\phi)R_{MN}=\widetilde{T}_{MN}-\frac{1}{3}g_{MN}\widetilde{T},
\end{eqnarray}
where $\widetilde{T}=g^{MN}\widetilde{T}_{MN}$ and
\begin{eqnarray}
 \widetilde{T}_{MN}&=& \partial_{M}\phi\partial_{N}\phi-
           g_{MN}\Big[\frac{1}{2}g^{PQ}\partial_{P}\phi\partial_{Q}\phi+V(\phi)\Big]
               \nonumber\\
              &&~~  +2\nabla_{M}\nabla_{N}F(\phi) -2 g_{MN}\Box F(\phi).
\end{eqnarray}
From the above
equations, we will obtain the linearized equations for the metric
fluctuations $h_{MN}$.

Under the axial gauge conditions $h_{5M}=0$, the total metric can be
written in the form
\begin{eqnarray}\label{MetricFlu}
 ds^{2}=\text{e}^{2A(y)}(\eta_{\mu\nu}+h_{\mu\nu})dx^{\mu}dx^{\nu}+dy^{2}.
\end{eqnarray}
The Ricci tensor can be computed from the metric (\ref{MetricFlu})
\begin{eqnarray}
 R_{MN}=R_{MN}^{(0)}+R_{MN}^{(1)}+\cdots,
\end{eqnarray}
where the zero order terms are
\begin{eqnarray}
 && R_{\mu\nu}^{(0)}=-\text{e}^{2A}(A''+4A'^{2})\eta_{\mu\nu}, \quad
 R_{55}^{(0)}=-4(A''+A'^{2}), \nonumber\\
 && R_{5\mu}^{(0)}=0,\nonumber
\end{eqnarray}
and the first order terms are
\begin{eqnarray}
 R_{\mu\nu}^{(1)}&=&-\text{e}^{2A}
            (\frac{1}{2}\partial_{y}^{2}+2A'\partial_{y}+A''+4A'^{2})h_{\mu\nu}
    \nonumber\\
          & &~  -\frac{1}{2}\Box^{(4)}h_{\mu\nu}
            -\frac{1}{2}\eta_{\mu\nu}\text{e}^{2A}A'
            \partial_{y}(\eta^{\alpha\beta}h_{\alpha\beta}),
            \nonumber\\
            & &~ -\frac{1}{2}\eta^{\alpha\beta} (\partial_{\mu}\partial_{\nu}h_{\alpha\beta}
                    -\partial_{\mu}\partial_{\alpha}h_{\nu\beta}
                    -\partial_{\nu}\partial_{\alpha}h_{\mu\beta}), ~~~~\\
 R_{55}^{(1)}&=&-\frac{1}{2}(\partial_{y}^{2}+2A'\partial_{y})\eta^{\mu\nu}h_{\mu\nu},\\
 R_{5\mu}^{(1)}&=&\frac{1}{2}\eta^{\alpha\beta}\partial_{y}(\partial_{\alpha}h_{\mu\beta}
                   -\partial_{\mu}h_{\alpha\beta}).
\end{eqnarray}
And we can also obtain
\begin{eqnarray}\label{Tmn2}
 \widetilde{T}_{MN}=\widetilde{T}_{MN}^{(0)}+\widetilde{T}_{MN}^{(1)}+\cdots,
\end{eqnarray}
with
\begin{eqnarray}\label{Tmn-0}
 \widetilde{T}_{\mu\nu}^{(0)}&=&
               -\text{e}^{2A}\eta_{\mu\nu}\Big[6\frac{dF}{d\phi}\phi'A'
           +2\frac{d^{2}F}{d\phi^{2}}\phi'^{2}
           \nonumber\\
           & &~~~~~~~~~~~~~
           +2\frac{dF}{d\phi}\phi''
                 +\frac{1}{2}\phi'^{2}+V(\phi)\Big],\\
 \widetilde{T}_{55}^{(0)}&=&\frac{1}{2}\phi'^{2}-V(\phi)-8\frac{dF}{d\phi}\phi'A',\quad
 \widetilde{T}_{5\mu}^{(0)}=0,
\end{eqnarray}
and
\begin{eqnarray}\label{Tmn-1}
 \widetilde{T}_{\mu\nu}^{(1)}&=&
               -\text{e}^{2A}h_{\mu\nu}\Big[6\frac{dF}{d\phi}\phi'A'
                 +2\frac{d^{2}F}{d\phi^{2}}\phi'^{2}+2\frac{dF}{d\phi}\phi''+\frac{1}{2}\phi'^{2}
                 \nonumber\\
           &&~
                 +V(\phi)\Big]
           +\text{e}^{2A}\frac{dF}{d\phi}\phi'h_{\mu\nu}'
                     -\text{e}^{2A}\eta_{\mu\nu}\frac{dF}{d\phi}\phi'h',\\
 \widetilde{T}_{55}^{(1)}&=&-\frac{dF}{d\phi}\phi'h',\quad
 \widetilde{T}_{5\mu}^{(1)}=0,
\end{eqnarray}
where $h=\eta^{\mu\nu}h_{\mu\nu}$. From above equations and
considering the fluctuations $h_{\mu\nu}$ satisfy the transverse
$\partial^{\mu}h_{\mu\nu}=0$ and traceless
$h=\eta^{\mu\nu}h_{\mu\nu}=0$ conditions, we can obtain the
following equations
\begin{eqnarray}\label{GravitonEQy}
 \left[\partial_{y}^{2}+Q'(y)\partial_{y}+\text{e}^{-2A}\Box^{(4)}\right]h_{\mu\nu}^{TT}(x,y)=0,
\end{eqnarray}
where we have set
\begin{eqnarray}
 Q(y)=4A(y)+{\ln}F(\phi(y)).
\end{eqnarray}
Hence, as we required in Sec.~\ref{SecModel}, $F(\phi(y))$ should
satisfy $F(\phi(y))>0$. By using the decomposition
$h_{\mu\nu}^{TT}(x,y)=\text{e}^{ipx}\tilde{h}_{\mu\nu}(y)$, the
above equation can be reexpressed as
\begin{eqnarray}
 \left[\partial_{y}^{2}+Q'(y)\partial_{y}+m^{2}\text{e}^{-2A}\right]\tilde{h}_{\mu\nu}(y)=0
\end{eqnarray}
with $p^{\mu}p_{\mu}=-m^{2}$ the four-dimensional mass of the
gravitons.

By using the conformally flat metric (\ref{line_z}), Eq.
(\ref{GravitonEQy}) can be rewritten as
\begin{eqnarray}\label{GravitonEQz}
 \left[\partial_{z}^{2} + \widetilde{Q}'(z)\partial_{z}+m^{2}
                \right]\tilde{h}_{\mu\nu}(z)=0,
\end{eqnarray}
where $\widetilde{Q}(z)=3A(z)+\ln F(\phi(z))$. By using the
transformation
$\tilde{h}_{\mu\nu}(z)=\text{e}^{-\widetilde{Q}/2}\varphi_{\mu\nu}(z)$,
Eq. (\ref{GravitonEQz}) can be expressed as the Schr\"{o}dinger
equation
\begin{eqnarray}\label{Schrodinger_eq}
 \left[-\partial_{z}^{2}+V_{\text{QM}}(z)\right]\varphi(z)=m^{2}\varphi(z),
\end{eqnarray}
where $\varphi_{\mu\nu}(z)$ has been replaced by $\varphi(z)$ for
simplicity and the localization potential is read as
\begin{eqnarray}\label{VQM}
  V_{\text{QM}}(z)=\frac{1}{2}\widetilde{Q}''(z)+\frac{1}{4}\widetilde{Q}'^{2}(z),
\end{eqnarray}
where the prime denotes the derivative with respect to $z$. Eq.
(\ref{Schrodinger_eq}) can be rewritten alternatively as
\begin{eqnarray}
 \bigg[\partial_{z}+\frac{1}{2}\widetilde{Q}'(z)\bigg]
 \bigg[ \partial_{z}-\frac{1}{2}\widetilde{Q}'(z)\bigg]
        \varphi(z)=-m^{2}\varphi(z),
\end{eqnarray}
so it will have a complete system of eigenstates with nonnegative
eigenvalues, i.e., $m^{2}\geq0$.

By setting $m=0$, the massless wave function (the zero mode) can be
obtained
\begin{eqnarray}\label{GravityZero}
 \varphi_{0}(z)\propto
 \text{e}^{\frac{3}{2}A(z)}F^{\frac{1}{2}}(\phi(z)).
\end{eqnarray}
If the zero mode satisfies the normalization condition
\begin{eqnarray}\label{norConGravz}
 \int_{-\infty}^{+\infty}\varphi_{0}^{2}(z)dz
    =\int_{-\infty}^{+\infty}\text{e}^{3A(z)}F(\phi(z))dz<\infty,
\end{eqnarray}
the zero mode is localized on the brane.

\begin{figure*}[htb]
\begin{center}
\subfigure[$k=1$]{\label{fig_VQMI}
\includegraphics[width=7cm]{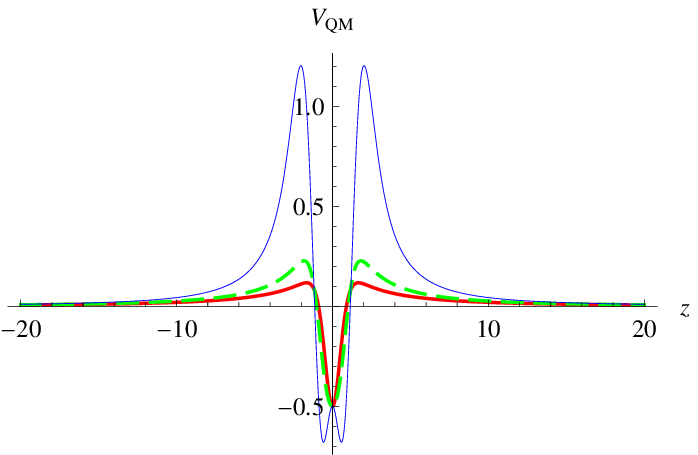}}
\subfigure[$k=1$]{\label{fig_ZeroGI}
\includegraphics[width=7cm]{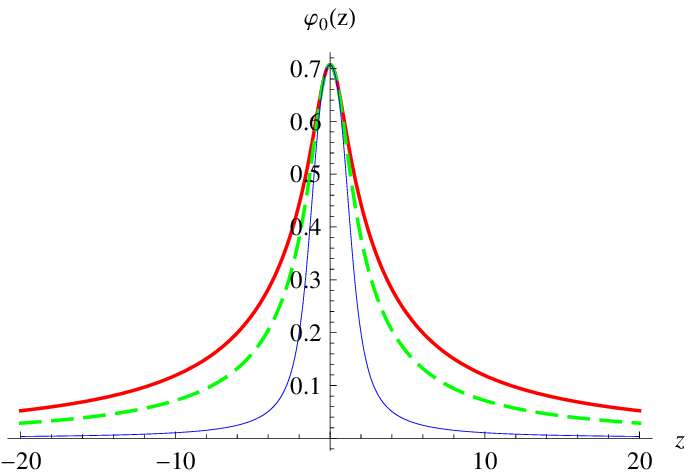}}
\end{center}\vskip -5mm
\caption{The shapes of the potential of gravity KK modes
$V_{\text{QM}}(z)$ and the gravity zero mode $\varphi_{0}(z)$. The
parameters are set to $\xi=0.1$ for the thick red line, $\xi=0.5$
for the dashed green line, and $\xi=0.9$ for the thin blue line. The
other parameters are set as $\phi_{0}=1$, $b=1$, and $k=1$. }
 \label{fig_VQMzeroI}
\end{figure*}

\begin{figure*}[htb]
\begin{center}
\subfigure[$k=3$]{\label{fig_VQM2}
\includegraphics[width=7cm]{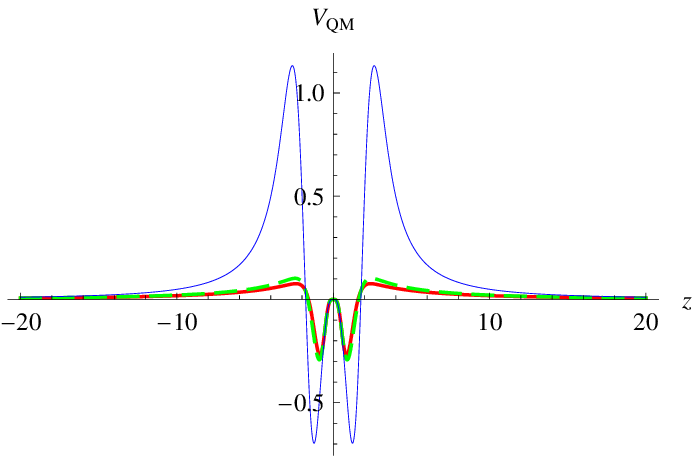}}
\subfigure[$k=3$]{\label{fig_ZeroG2}
\includegraphics[width=7cm]{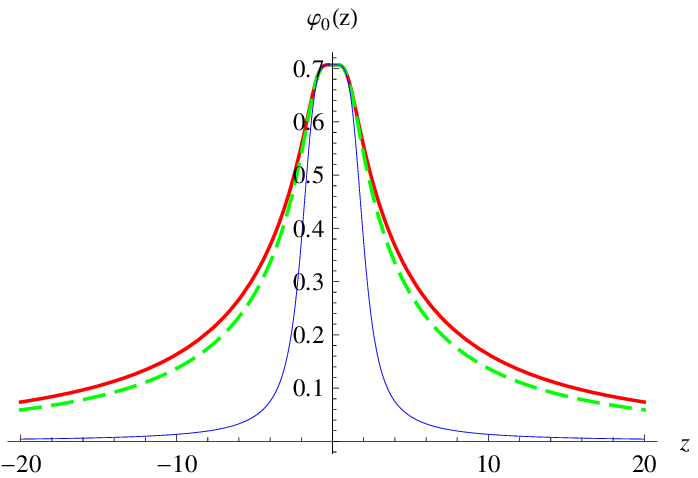}}
\end{center}\vskip -5mm
\caption{The shapes of the potential of gravity KK modes
$V_{\text{QM}}(z)$ and the gravity zero mode $\varphi_{0}(z)$. The
parameters are set to $\xi=0.1$ for the thick red line, $\xi=0.3$
for the dashed green line, and $\xi=0.9$ for the thin blue line. The
other parameters are set as $\phi_{0}=1$, $b=1$, and $k=3$. }
 \label{fig_VQMzeroII}
\end{figure*}

From the solutions which we have shown in Sec.~\ref{SecModel}, the
potential of gravity KK modes $V_{\text{QM}}$ (\ref{VQM}) and the
gravity zero mode (\ref{GravityZero}) are also can be obtained,
which are shown in Figs.~\ref{fig_VQMzeroI} and \ref{fig_VQMzeroII}.

By considering the asymptotic behavior of the warp factor and the
background scalar, the asymptotic behavior of the potential as
$z\rightarrow\pm\infty$ is
\begin{eqnarray}
 V_{\text{QM}}(z\rightarrow\pm\infty)\rightarrow 0,
\end{eqnarray}
which is similar to a volcano-type potential. The asymptotic
behavior of the gravity zero mode is
\begin{eqnarray}
 \varphi_{0}(z\rightarrow\pm\infty)
         \propto\frac{1}{2}(1-\xi\phi_{0}^{2})^{\frac{1}{2}}
           \bigg(\frac{1}{b|z|}\bigg)^{\frac{3}{2}}.
\end{eqnarray}
and the normalization condition (\ref{norConGravz}) is equivalent to
the following condition
\begin{eqnarray}
 \int_{1}^{+\infty}\frac{1}{4}(1-\xi\phi_{0}^{2})
             (\frac{1}{b|z|})^{3}dz
             = \frac{1-\xi\phi_{0}^{2}}{8b^{3}}
        <\infty,
\end{eqnarray}
so it is clear that the normalization condition is satisfied. For
$k=1$, when $\xi$ is smaller than the critical coupling constant
$\xi_{0}$, the zero mode is localized on the single brane, and when
$\xi$ is close to $\xi_0$ or $\xi>\xi_{0}$, the zero mode is
localized at the center of the two sub-branes. For $k=3$, the zero
mode is localized between the two sub-branes. The massive gravity KK
modes are the continuous spectrum wave functions with $m^{2}>0$, and
asymptotically turn into plane waves.

Generally, the volcano-type potential implies that there may exist
resonant states, which are quasilocalized on the branes. So we will
study whether the resonant states exist in this system. From Figs.
\ref{fig_VQMzeroI} and \ref{fig_VQMzeroII}, it is seen that with the
coupling constant $\xi$ becoming larger, the height of the potential
$V_{\text{QM}}$ gets higher, with which the resonant states could be
obtained more easily. When the coupling constant $\xi=0.9$ and other
parameters are set as $\phi_{0}=1$ and $b=1$, there is only one
resonant state for the double-kink background scalar field. However,
we find that when the coupling constant $\xi=0.99$ and other
parameters are set as $\phi_{0}=1$ and $b=1$, there is a resonant KK
mode for the single-kink background scalar field, and two resonant
KK modes for the double-kink background scalar field. Because
resonant states are oscillating when far away from the brane along
the extra dimension, they can not be normalized. As Ref.
\cite{Liu0907.0910}, we use the function
\begin{equation}
 P_{\text{G}}(m^{2})=\frac{\int_{-z_b}^{z_b} |\varphi(z)|^2 dz}
                 {\int_{-z_{max}}^{z_{max}} |\varphi(z)|^2 dz}
 \label{ProbabilityGravity}
\end{equation}
to describe the relative probability for finding the resonances on a
thick brane. Here $2z_{b}$ is about the width of the thick brane and
$z_{max}$ is set to $z_{max}=10z_{b}$. It is clear that for KK modes
with $m^{2}\gg V_{\text{QM}}^{\text{max}}$
($V_{\text{QM}}^{\text{max}}$ is the maximum value of
$V_{\text{QM}}$), which are approximately taken as plane waves, the
value of $P_{\text{G}}(m^{2})$ will trend to $\frac{1}{10}$.

\begin{figure*}[htb]
\begin{center}
\subfigure[$\xi=0.9,~k=1$]{\label{fig_PGka}
\includegraphics[width=7cm]{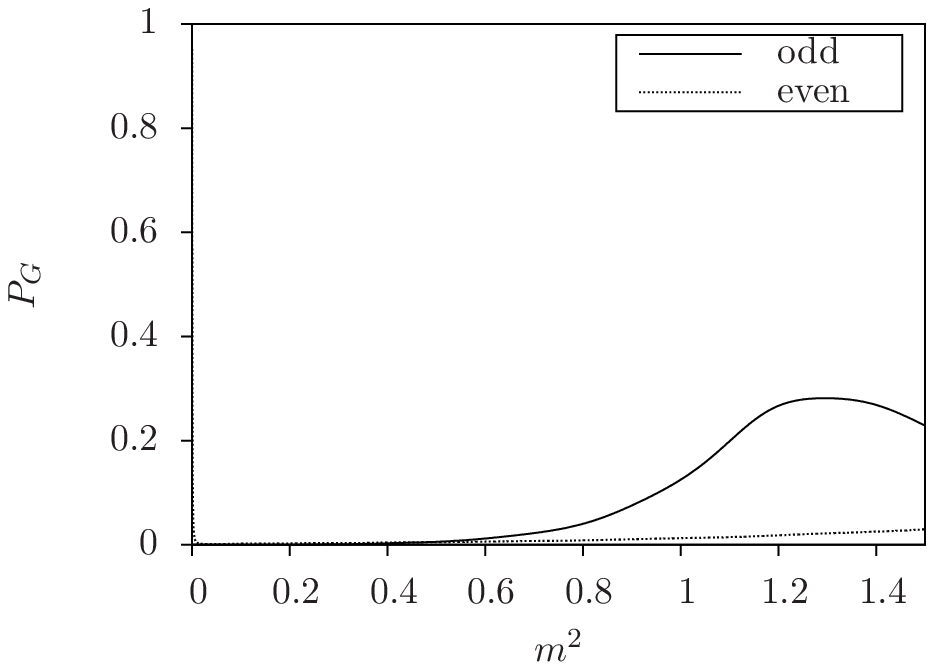}}
\subfigure[$\xi=0.9,~k=3$]{\label{fig_PGka}
\includegraphics[width=7cm]{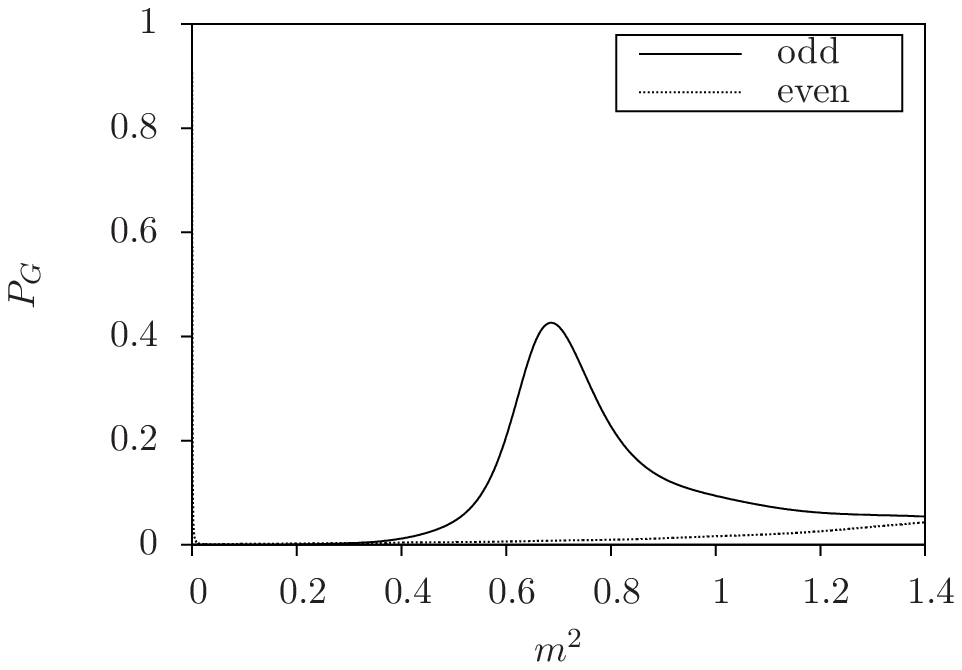}}
\end{center}\vskip -5mm
\caption{ The profiles of $P_{\text{G}}$ for massive even and odd
gravity KK modes for the parameters $\xi=0.9$, $\phi_{0}=1$, $b=1$,
$k=1$ and $k=3$. }
 \label{fig_PGk3a}
\end{figure*}

\begin{figure*}[htb]
\begin{center}
\subfigure[$\xi=0.99,~k=1$]{\label{fig_PG1}
\includegraphics[width=7cm]{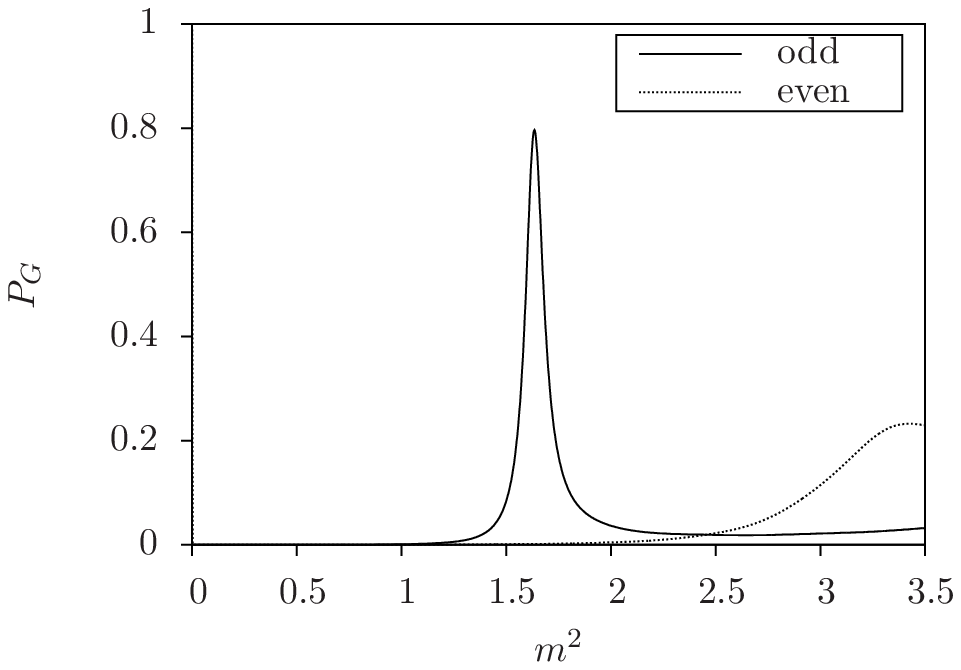}}
\subfigure[$\xi=0.99,~k=3$]{\label{fig_PG2}
\includegraphics[width=7cm]{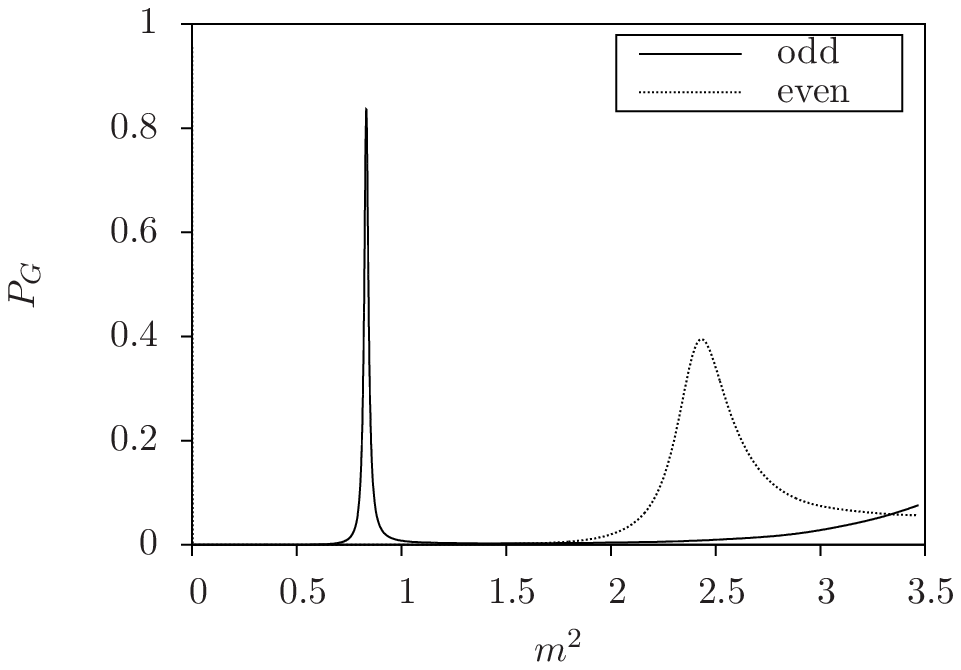}}
\end{center}\vskip -5mm
\caption{ The profiles of $P_{\text{G}}$ for massive even and odd
gravity KK modes for the parameters $\xi=0.99$, $\phi_{0}=1$, $b=1$,
$k=1$ and $k=3$. }
 \label{fig_PG}
\end{figure*}

\begin{figure*}[htb]
\begin{center}
\subfigure[$\xi=0.9,~k=3$,~$n=1$]{
\includegraphics[width=7cm]{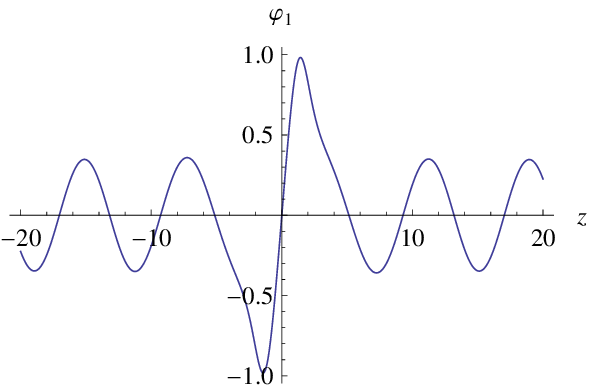}}
\end{center}\vskip -5mm
\caption{ The wave function of the gravity resonant KK mode for the
double-kink background scalar. The parameters are set to $\xi=0.9$,
$\phi_{0}=1$, $b=1$ and $k=3$. }
 \label{fig_GResk3a}
\end{figure*}

\begin{figure*}[htb]
\begin{center}
\subfigure[$\xi=0.99,~k=1$,~$n=1$]{
\includegraphics[width=7cm]{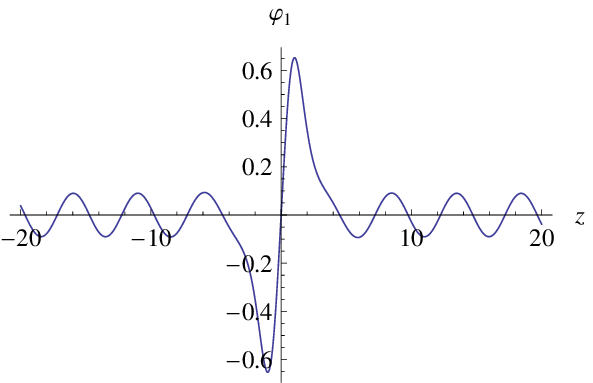}}
\end{center}\vskip -5mm
\caption{ The wave function of the gravity resonant KK mode for the
single-kink background scalar. The parameters are set to $\xi=0.99$,
$\phi_{0}=1$, $b=1$ and $k=1$. }
 \label{fig_GResone}
\end{figure*}

\begin{figure*}[htb]
\begin{center}
\subfigure[$\xi=0.99,~k=3$,~$n=1$]{\label{fig_PG1}
\includegraphics[width=7cm]{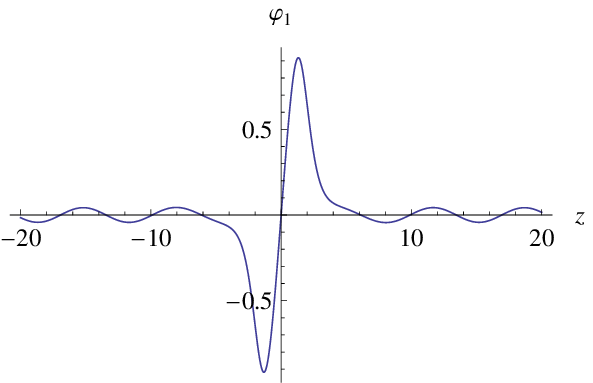}}
\subfigure[$\xi=0.99,~k=3$,~$n=2$]{\label{fig_PG2}
\includegraphics[width=7cm]{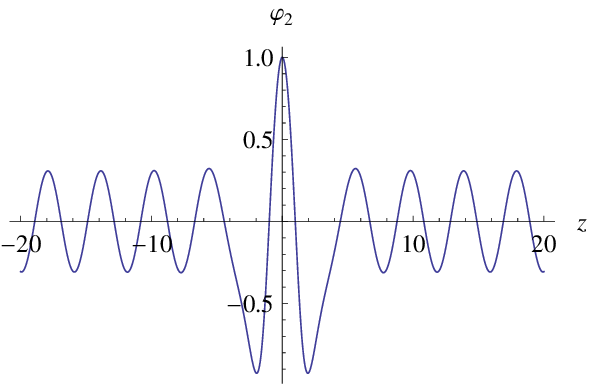}}
\end{center}\vskip -5mm
\caption{ The wave functions of the gravity resonant KK modes for
the double-kink background scalar. The parameters are $\xi=0.99$,
$\phi_{0}=1$, $b=1$ and $k=3$. }
 \label{fig_GRestwo}
\end{figure*}

The profiles of $P_{\text{G}}$ corresponding to the coupling
constant $\xi=0.9$ for single-kink scalar field $(k=1)$ and
double-kink background scalar field $(k=3)$ are shown in
Fig.~\ref{fig_PGk3a} for the thick brane, and $P_{\text{G}}$
corresponding to $\xi=0.99$ for $k=1$ and $k=3$ are shown in
Fig.~\ref{fig_PG}. In these figures, each peak corresponds to a
resonant state. When $\xi=0.9$, it is seen that only for the
double-kink background scalar field there is a resonant KK mode.
When $\xi=0.99$, there are one and two resonant KK modes for
single-kink and double-kink background scalars, respectively. Hence,
we come to the conclusion that if the coupling constant $\xi$
becomes larger, the number of the resonant gravity KK modes will
increase. As Ref. \cite{PRL845928}, we estimate the lifetime $\tau$
of a resonant state as $\tau\sim \Gamma^{-1}$ with $\Gamma=\delta m$
being the full width at half maximum of the peak. Then the mass $m$,
width $\Gamma$ and lifetime $\tau$ of the resonant KK modes are
listed in Table~\ref{tab1}, and the resonant KK modes are shown in
Figs.~ \ref{fig_GResk3a}, \ref{fig_GResone} and \ref{fig_GRestwo}.

\begin{table*}[h]
\begin{center}
\renewcommand\arraystretch{1.3}
\begin{tabular}
 {|l|c|c|c|c|c|c|c|}
  \hline
 ~~~$\xi$~ & $~k~$ & ~Height of $V_{\text{QM}}$~ & ~$n$~ & ~$m^{2}$~ & ~$m$~ & ~$\Gamma$~ & ~$\tau$~   \\
    \hline\hline
 $0.9$  & $3$ & $V_{\text{QM}}^{\text{max}}=1.1322$
     & ~$1$~ &~0.685649 ~& ~0.82804~ & ~0.123867~ & ~8.07321~ \\
   \hline\hline
 $0.99$  & $1$ & $V_{\text{QM}}^{\text{max}}=2.8051$
     & ~$1$~ &~1.634424 ~& ~1.27845~ & ~0.041632~ & ~24.02015~ \\
   \cline{2-8}
    & $3$ & $V_{\text{QM}}^{\text{max}}=2.7746$
       & ~$1$~ & ~0.831835~ & ~0.91205~ & ~0.013373~ & ~74.77949~  \\
  \cline{4-8}
  &   &
     & ~$2$~ & ~2.433646~ & ~1.56001~ & ~0.104536~ & ~9.56606~   \\
  \hline
\end{tabular}
\end{center}
\caption{The mass, width, and lifetime of the resonances of gravity.
The parameters are set to $\phi_{0}=1$ and $b=1$. Here $n$ is the
order of the resonant states with corresponding $m^2$ from small to
large.} \label{tab1}
\end{table*}

\section{Localization of the various matters on the thick branes}
\label{SecLocalization}

In this section, we will investigate the effect of nonminimally
coupled parameter $\xi$ on the localization and resonances of
various bulk matter fields on the thick branes, by presenting the
potential of the corresponding Schr\"{o}dinger equation for the KK
modes of the various matter fields. Spin-0 scalars, spin-1 vectors
and spin-1/2 fermions will be considered by means of the
gravitational interaction.

\subsection{Scalar fields and vector fields}

We first consider the localization of scalar and vector fields on
the thick branes obtained in the previous section, then turn to
fermions in the next subsections. Let us start by considering the
action of a massless real scalar coupled to gravity
\begin{eqnarray}\label{action_scalar}
S_{0}=-\frac{1}{2}\int d^{5}x\sqrt{-g}~
          g^{MN}\partial_{M}\Phi\partial_{N}\Phi.
\end{eqnarray}
By using of the KK decomposition
$\Phi(x,z)=\sum_{n}\phi_{n}(x)\chi_{n}(z)\text{e}^{-3A/2}$, it is
easy to derive the equations for the scalar KK modes:
\begin{eqnarray}
 \left[-\partial^{2}_z+ V_{0}(z)\right]{\chi}_n(z)
           =m_{n}^{2} {\chi}_{n}(z),
  \label{SchEqScalar1}
\end{eqnarray}
where the Schr\"{o}dinger potential is given by
\begin{eqnarray}\label{VScalar}
  V_0(z)=\frac{3}{2} A''(z) + \frac{9}{4}A'^{2}(z).
\end{eqnarray}
Here $m_{n}$ are the masses of the scalar KK modes, they are also
the masses of the four-dimensional scalars $\phi_{n}(x)$. It is
clear that the potential $V_{0}(z)$ defined in (\ref{VScalar}) is a
four-dimensional mass-independent potential. Note that those scalar
KK modes localized on the branes should satisfies the following
orthonormality conditions:
\begin{eqnarray}
 \int^{+\infty}_{-\infty}
 \;\chi_m(z)\chi_n(z) dz=\delta_{mn}.
 \label{normalizationCondition1}
\end{eqnarray}

\begin{figure*}[htb]
\begin{center}
\subfigure[$k=1$]{\label{fig_VScalarI}
\includegraphics[width=7cm]{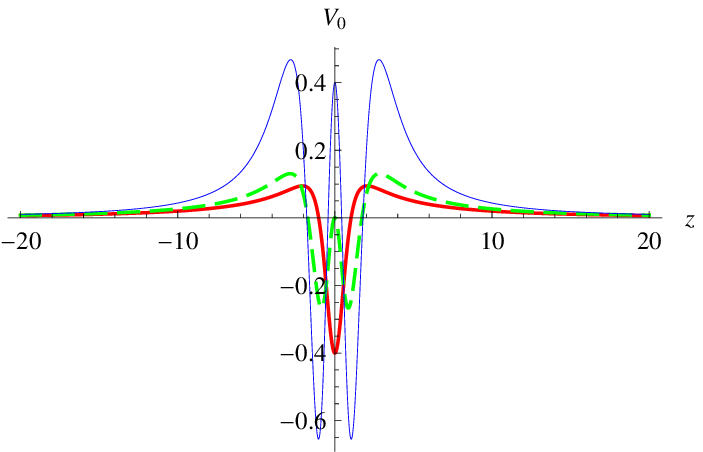}}
\subfigure[$k=1$]{\label{fig_zeroScalarI}
\includegraphics[width=7cm]{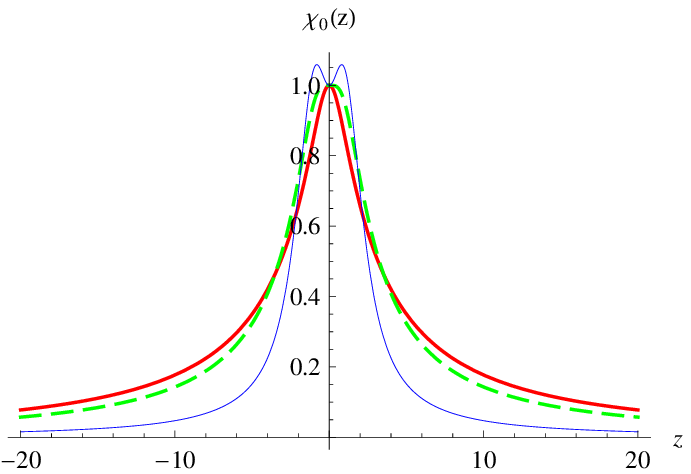}}
\end{center}\vskip -5mm
\caption{The shapes for the potential of the scalar KK modes
$V_{0}(z)$ and the scalar zero mode $\chi_{0}(z)$. The parameters
are set to $\xi=0.1$ for the thick red line, $\xi=0.5$ for the
dashed green line, and $\xi=0.9$ for the thin blue line. The other
parameters are set as $\phi_{0}=1$, $b=1$, and $k=1$. }
 \label{fig_ScalarI}
\end{figure*}

\begin{figure*}[htb]
\begin{center}
\subfigure[$k=3$]{\label{fig_VScalarII}
\includegraphics[width=7cm]{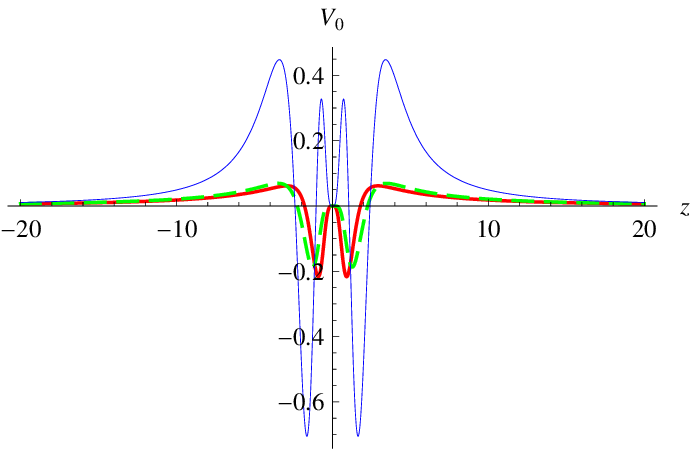}}
\subfigure[$k=3$]{\label{fig_zeroScalarII}
\includegraphics[width=7cm]{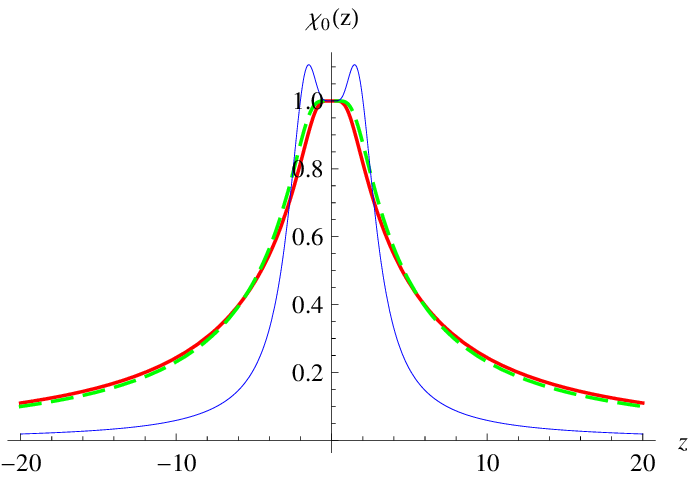}}
\end{center}\vskip -5mm
\caption{The shapes for the potential of the scalar KK modes
$V_{0}(z)$ and the scalar zero mode $\chi_{0}(z)$. The parameters
are set to $\xi=0.1$ for the thick red line, $\xi=0.3$ for the
dashed green line, and $\xi=0.9$ for the thin blue line. The other
parameters are set as $\phi_{0}=1$, $b=1$, and $k=3$. }
 \label{fig_ScalarII}
\end{figure*}

For the thick branes considered previous section, the warp factor
can not be written as an analytical form, however, the potential
$V_{0}$ can be solved numerically, which is shown in
Figs.~\ref{fig_ScalarI} and \ref{fig_ScalarII}.

The scalar zero mode can be solved from (\ref{SchEqScalar1}) by
setting $m_{0}=0$:
\begin{eqnarray}
 \chi_{0}\propto \text{e}^{\frac{3}{2}A(z)}. \label{Scalar0Mode}
\end{eqnarray}
This scalar zero mode (hence the four-dimensional massless scalar)
is localized on the branes because the boundaries of the
five-dimensional space-time along extra dimension are AdS. This can
also be confirmed by the following simple calculation:
\begin{eqnarray}
 \chi_{0}(z\rightarrow\pm\infty)\rightarrow\bigg(\frac{1}{b|z|}\bigg)^{\frac{3}{2}},\\
  \int_{1}^{\infty}
             \bigg(\frac{1}{b|z|}\bigg)^{3}dz=\frac{1}{2b^3}
        <\infty.
\end{eqnarray}
In fact, this is a well known conclusion in brane-world model. Here,
our point is another question: where the four-dimensional massless
scalar locates when the thick branes split? As far as we know, when
a thick brane splits into two sub-branes in minimally coupled
theories, the four-dimensional massless scalar is located between
the two sub-branes. Here, for the case of nonminimally coupled
theory considered in this papaer, we can see from Figs.
\ref{fig_ScalarI} and \ref{fig_ScalarII}, as well as Eqs.
(\ref{Scalar0Mode}), (\ref{Azk1}) and (\ref{Azk3}) that, the
four-dimensional massless scalar is localized at the center of the
two sub-branes generated by single-kink or between them by
double-kink as $\xi<\xi_0$, just as the usual situation, while it is
localized on each sub-brane when the coupling parameter $\xi$ is
greater than its critical value $\xi_0$.

All other scalar KK modes are continuum and massive, for which we do
not find resonant KK modes in the spectrum. This is different the
case of gravity.

For Abelian spin-1 vectors described by the following
action
\begin{eqnarray}\label{action_Vector}
S_{1} = -\frac{1}{4}\int d^{5}x \sqrt{-g}~ g^{M N}
 g^{RS}F_{MR}F_{NS},
\end{eqnarray}
because the asymptotical form of the warp factor is same as the RS
brane-world when far away from the branes, with the same case of RS
brane the vector zero mode can not localized on the branes with AdS
boundaries. This problem can be solve in some higher dimensional
models \cite{OdaPLB2000113} or in thick dS branes and Weyl thick
branes models \cite{Liu0708}. We do not yet find vector resonant KK
modes in the spectrum.

\subsection{Spin-1/2 fermion fields}

Next, we will study the localization and resonances of fermions on
the thick branes. In five-dimensional space-time, fermions are
four-component spinors and their Dirac structure can be described by
the curved space gamma matrices
$\Gamma^{M}=e^{-A}(\gamma^{\mu},\gamma^{5})$, where $\gamma^{\mu}$
and $\gamma^{5}$ are the usual flat gamma matrices in the
four-dimensional Dirac representation. The Dirac action of a
massless spin-1/2 fermion coupled to the background scalar $\phi$
(\ref{BackgroundScalar}) is
\begin{eqnarray}
 S_{1/2} = \int d^5 x \sqrt{-g} \left(\bar{\Psi} \Gamma^M
          D_{\mu} \Psi
          -\eta \bar{\Psi} G(\phi)\Psi\right),
\label{DiracAction}
\end{eqnarray}
where $\eta$ is a coupling constant, and
$D_{\mu}\Psi=(\partial_M+\omega_M) \Psi$ with the nonvanishing
components of the spin connection $\omega_M$ for the background
metric (\ref{line_z}) given by $\omega_\mu =\frac{1}{2}A' \gamma_\mu
\gamma_5.$ Then the equation of motion reads
\begin{eqnarray}
 \big[ \gamma^{\mu}\partial_{\mu}
         + \gamma^5 \left(\partial_z  +2 A'\right)
         -\eta\; \text{e}^A G(\phi)
 \big] \Psi =0, \label{DiracEq1}
\end{eqnarray}
where $\gamma^{\mu}\partial_{\mu}$ is Dirac operator on the brane.

Now we would like to investigate the effect of the coupling
parameter $\xi$ on the localization and resonances of the Dirac
spinor on the branes by studying the above five-dimensional Dirac
equation. Because of the Dirac structure of the fifth gamma matrix
$\gamma^{5}$, we expect that the left- and right-chiral projections
of the four-dimensional part have different behaviors. From the
equation of motion (\ref{DiracEq1}), we will search for the
solutions of the general chiral decomposition
\begin{equation}\label{FerKKdecom}
 \Psi = \sum_n\big[\psi_{L,n}(x) L_{n}(z)
 +\psi_{R,n}(x) R_{n}(z)\big]\text{e}^{-2A},
\end{equation}
where $\psi_{L}=\frac{1-\gamma^5}{2}\psi$ and
$\psi_{R}=\frac{1+\gamma^5}{2} \psi$ are the left- and right-chiral
components of a four-dimensional Dirac field $\psi$, respectively.
By demanding $\psi_{L,R}$ satisfy the four-dimensional massive Dirac
equations $\gamma^{\mu}(\partial_{\mu}+\hat{\omega}_\mu)\psi_{L,R}
=m\psi_{R,L}$, we obtain the following Schr\"{o}dinger-like
equations for the KK modes of the left- and right-chiral fermions:
\begin{subequations}\label{SchEqFermion}
\begin{eqnarray}
  \big(-\partial^2_z + V_L(z) \big)L_{n}
            &=&m^2 L_{n},~~
   \label{SchEqLeftFermion}  \\
  \big(-\partial^2_z + V_R(z) \big)R_{n}
            &=&m^2 R_{n},
   \label{SchEqRightFermion}
\end{eqnarray}
\end{subequations}
where the effective potentials are
\begin{subequations}\label{Vfermion}
\begin{eqnarray}
  V_L(z)&=& \big(\eta \text{e}^{A} G(\phi)\big)^2
     -\partial_z \big(\eta \text{e}^{A} G(\phi)\big), \label{VL}\\
  V_R(z)&=&   V_L(z)|_{\eta \rightarrow -\eta}. \label{VR}
\end{eqnarray}
\end{subequations}

For the purpose of getting the standard four-dimensional action for
a massless fermion and an infinite sum of the massive fermions:
\begin{eqnarray}
 S_{\frac{1}{2}} &=& \int d^5 x \sqrt{-g} ~\bar{\Psi}
     \left[  \Gamma^M (\partial_M+\omega_M)
     -\eta G(\phi)\right] \Psi  \nonumber \\
  &=&\sum_{n}\int d^4 x \sqrt{-\hat{g}}
    ~\bar{\psi}_{n}
      [\gamma^{\mu}(\partial_{\mu}+\hat{\omega}_\mu)
        -m_{n}]\psi_{n},~~~
\end{eqnarray}
we need the following orthonormality conditions for $L_n$ and $R_n$:
\begin{eqnarray}
 \int_{-\infty}^{\infty} L_m L_ndz
   &=& \delta_{mn}, \label{orthonormalityFermionL} \\
 \int_{-\infty}^{\infty} R_m R_ndz
   &=& \delta_{mn}, \label{orthonormalityFermionR}\\
 \int_{-\infty}^{\infty} L_m R_ndz
   &=& 0. \label{orthonormalityFermionR}
\end{eqnarray}

\begin{figure*}[htb]
\begin{center}
\subfigure[$k=1$]{\label{fig_VfLI}
\includegraphics[width=7cm]{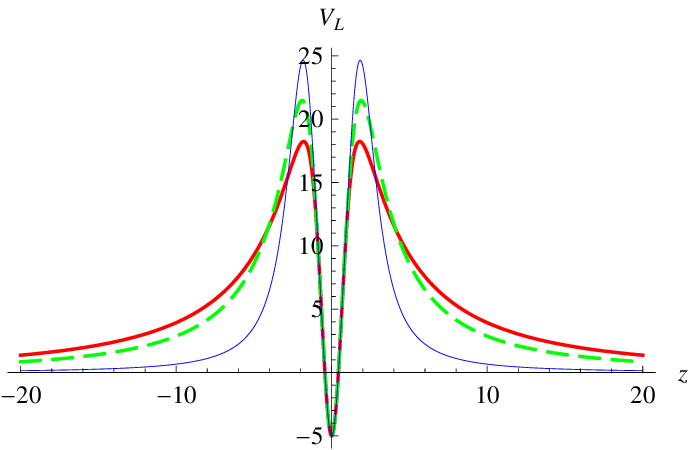}}
\subfigure[$k=1$]{\label{fig_VfRI}
\includegraphics[width=7cm]{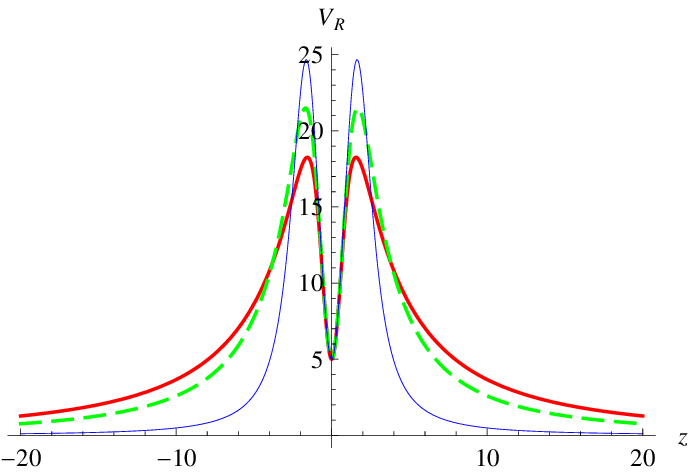}}
\end{center}\vskip -5mm
\caption{The shapes of the potentials for the left- and right-chiral
fermions coupled with single-kink scalar ($k=1$). The parameters are
set to $\xi=0.1$ for the thick red line, $\xi=0.5$ for the dashed
green line, and $\xi=0.9$ for the thin blue line. The other
parameters are set to $\phi_{0}=1$, $\eta=5$, and $b=1$. }
 \label{fig_VfermionI}
\end{figure*}

\begin{figure*}[htb]
\begin{center}
\subfigure[$k=3$]{\label{fig_VfLII}
\includegraphics[width=7cm]{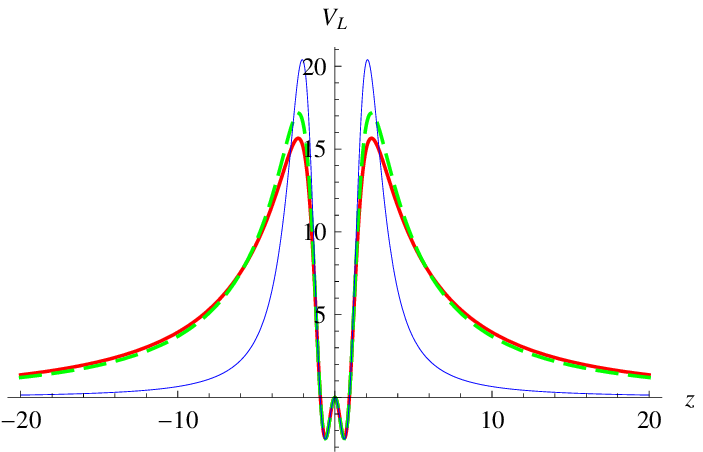}}
\subfigure[$k=3$]{\label{fig_VfRII}
\includegraphics[width=7cm]{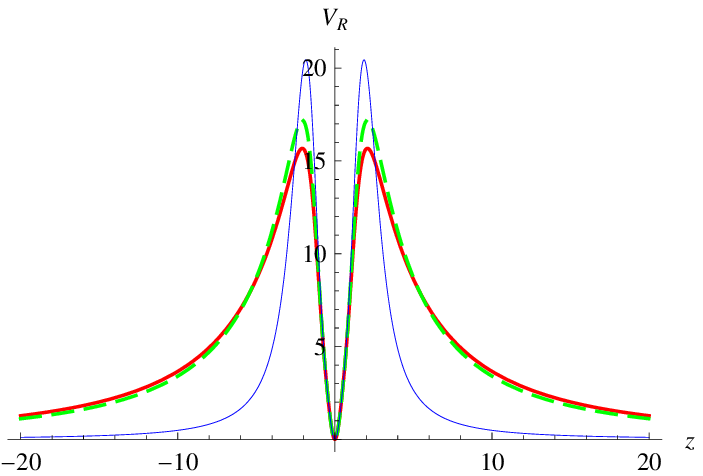}}
\end{center}\vskip -5mm
\caption{The shapes of the potentials for the left- and right-chiral
fermions coupled with double-kink scalar ($k=3$). The parameters are
set to $\xi=0.1$ for the thick red line, $\xi=0.3$ for the dashed
green line, and $\xi=0.9$ for the thin blue line. The other
parameters are set to $\phi_{0}=1$, $\eta=5$, and $b=1$. }
 \label{fig_VfermionII}
\end{figure*}

From Eqs. (\ref{SchEqFermion})and (\ref{Vfermion}), it is clear
that, in order to localize the left- and right-chiral fermions, some
kind of scalar-fermion coupling must be introduced. This situation
is similar to the one in Refs.
\cite{BajcPLB2000,OdaPLB2000113,Liu0708,Volkas0705.1584,Ringeval,Liu0907.0910,zhao_CQG},
in which the authors introduced the scalar-fermion coupling term
$m\bar{\Psi}F(\phi)\Psi$ for the localization of the fermion fields
on a brane. Moreover, if we demand that $V_{L}(z)$ and $V_{R}(z)$
are $Z_{2}$-even with respect to the extra dimension coordinate $z$,
$G(\phi)$ should be an odd function of the kink $\phi(z)$. In this
paper, we choose the simplest Yukawa coupling: $G(\phi)=\phi$. So
the potentials for left- and right-chiral fermion KK modes can be
expression as
\begin{eqnarray}
 V_{L}(z)&=& \eta^{2}\text{e}^{2A(z)}\phi^{2}(z)
             -\eta\text{e}^{A}\phi'(z)-\eta\phi(z)\text{e}^{A(z)}A'(z),
             ~~~~~\\
 V_{R}(z)&=& \eta^{2}\text{e}^{2A(z)}\phi^{2}(z)
             +\eta\text{e}^{A}\phi'(z)+\eta\phi(z)\text{e}^{A(z)}A'(z).
\end{eqnarray}
First, by considering the initial conditions of the warp factor and
the background scalar at $z=0$, the values of the potentials can be
obtained:
\begin{subequations}
\begin{eqnarray}
 V_{L}(0)&=&\left. -\eta\phi'(z)\right|_{z=0}, \\
 V_{R}(0)&=&\left. +\eta\phi'(z)\right|_{z=0},
\end{eqnarray}
\end{subequations}
So for $k=1$,
\begin{eqnarray}
 V_{L}(0)=-\eta\phi_0 b,  \quad\quad
 V_{R}(0)=+\eta\phi_0 b,
\end{eqnarray}
and for $k=3$,
\begin{eqnarray}
 V_{L,R}(0)=0.
\end{eqnarray}
Second, when far away from the brane $z\rightarrow\pm\infty$,
\begin{eqnarray}
 V_{L,R}(z\rightarrow\pm\infty)\rightarrow 0.
\end{eqnarray}
Here the potentials, which are
shown in Figs.~\ref{fig_VfermionI} and \ref{fig_VfermionII}, can be solved by the numerical method.

\begin{figure*}[htb]
\begin{center}
\subfigure[$k=1$]{\label{fig_VfLII}
\includegraphics[width=7cm]{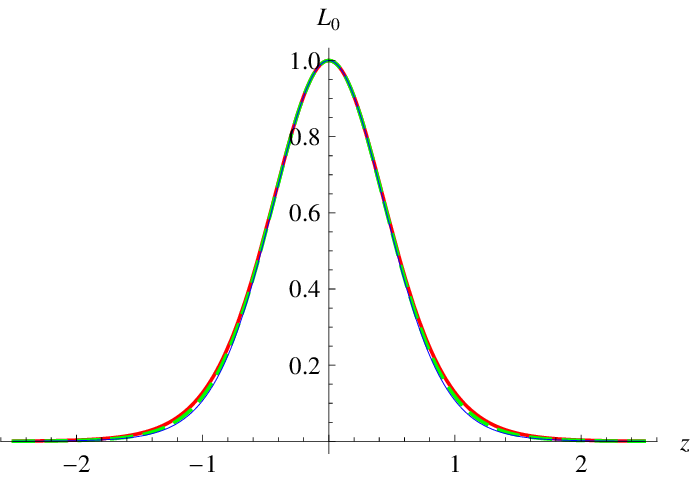}}
\subfigure[$k=3$]{\label{fig_VfRII}
\includegraphics[width=7cm]{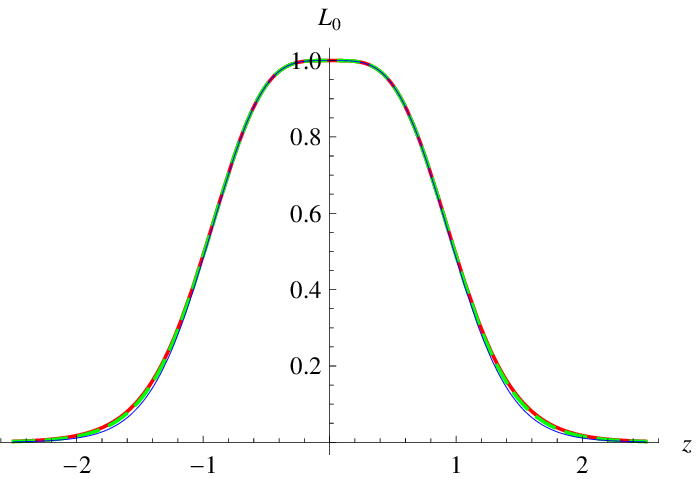}}
\end{center}\vskip -5mm
\caption{The shapes of the left-chiral fermion KK zero mode. The
parameters are set to $\xi=0.1$ for the thick red line, $\xi=0.5$
for the dashed green line, and $\xi=0.9$ for the thin blue line for
the left figure ($k=1$), and the parameters are set to $\xi=0.1$ for
the thick red line, $\xi=0.3$ for the dashed green line, and
$\xi=0.9$ for the thin blue line for the right figure ($k=3$). The
other parameters are set to $\phi_{0}=1$, $\eta=5$, and $b=1$. }
 \label{fig_LZero}
\end{figure*}

The left- and right-chiral fermion zero modes are solved as
\begin{subequations}
\begin{eqnarray}
 L_{0}(z)&\propto& \exp\left(-\eta\int^{z}_{0}dz' \text{e}^{A(z')}\phi(z')
 \right), \\
 R_{0}(z)&\propto& L_{0}(z)|_{\eta\rightarrow -\eta}.
\end{eqnarray}
\end{subequations}
From the asymptotic behavior of the warp factor as
$z\rightarrow\pm\infty$, the asymptotic behavior of left- and
right-chiral fermion zero mode can be analyzed:
\begin{subequations}
\begin{eqnarray}
 &&~L_{0}(z\rightarrow\pm\infty)\rightarrow
              |z|^{-\frac{\eta\phi_{0}}{b}}, \\
 &&~R_{0}(z\rightarrow\pm\infty)\rightarrow
              |z|^{ \frac{\eta\phi_{0}}{b}}.
\end{eqnarray}
\end{subequations}
So for the positive coupling constant $\eta$, only the left-chiral
zero mode tends to zero when far away from the branes, which may be
localized on the brane. We need further check whether the
normalization condition (\ref{orthonormalityFermionL}) is satisfied
for the left-chiral zero mode, i.e.,
\begin{eqnarray}
 \int_{-\infty}^{+\infty} dz L_{0}^{2}(z)<\infty.
\end{eqnarray}
Since the values of the zero modes are finite at finite $z$, the
above normalization condition is equivalent to the following
condition
\begin{eqnarray}
 \int_{1}^{\infty} dz |z|^{-\frac{2\eta\phi_{0}}{b}}<\infty.
\end{eqnarray}
Only when $\eta>\eta_{0}=\frac{b}{2\phi_{0}}$ (note that $\phi_{0}$
and $b$ are positive real parameters), the above integral is
convergent, which means that the left-chiral zero mode can be
localized on the branes under this condition. From the shapes of the
zero mode of left-chiral fermions (Fig.~\ref{fig_LZero}), it can be
seen that the zero mode is localized on the center of the two
sub-branes when the thick brane splits (single-kink case) or between
them (double-kink case). However, we will show in the following that
the nonminimally coupling parameter $\xi$ would effect the
resonance spectrum of both the left- and right-chiral fermions.

All the massive KK modes of the left- and right-chiral fermions are
the continuum modes, and can not be localized on the brane. Next, we
are going to investigate the quasilocalization of the left- and
right-chiral fermion KK modes, which are called resonances. As Sec.
\ref{SecGraviton}, the relative probability for finding the
resonances on a brane can be given as
\begin{eqnarray}
P_{\text{L,R}}(m^2)=\frac{\int^{z_b}_{-z_b}|L,R(z)|^{2}dz}
              {\int_{-z_{max}}^{z_{max}} |L,R(z)|^2 dz}.
\end{eqnarray}
In order to investigate the effect of the gravity-scalar coupling
constant $\xi$ on fermion resonances, we will investigate the
resonant KK modes with different values of $\xi$.

\begin{figure*}[htb]
\begin{center}
\subfigure[$\xi=0.9,~k=1,$~Left-chiral]{\label{fig_PLk1c}
\includegraphics[width=7cm]{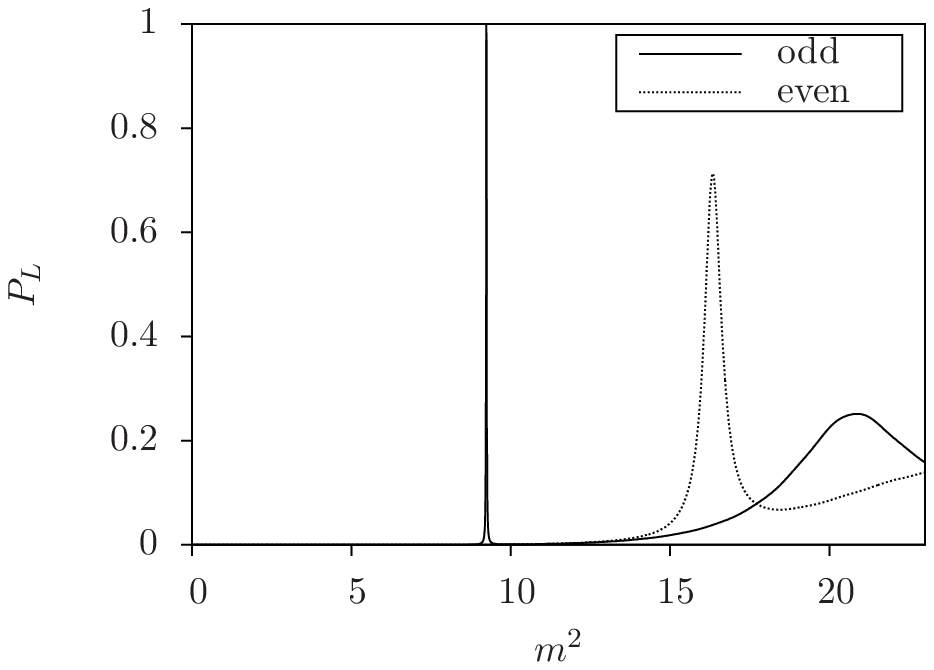}}
\subfigure[$\xi=0.9,~k=1,$~Right-chiral]{\label{fig_PRk1c}
\includegraphics[width=7cm]{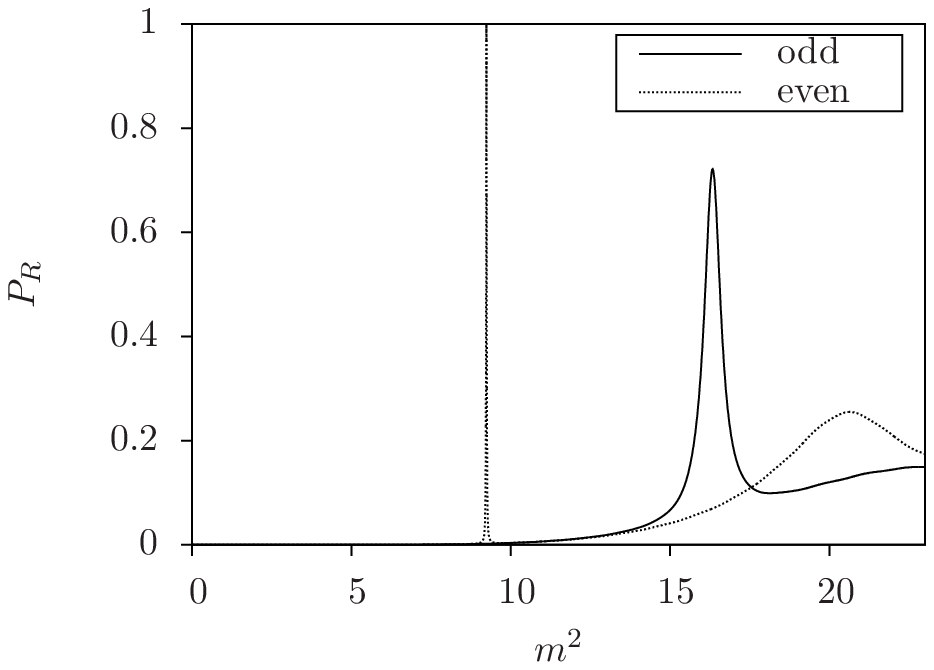}}
\end{center}\vskip -5mm
\caption{ The profiles of $P_{\text{L,R}}$ for the left- and
right-chiral fermion resonances for the brane with the parameters $k=1$,
$\xi=0.9$, $\eta=5$, $\phi_{0}=1$ and $b=1$. }
 \label{fig_Pfk1c}
\end{figure*}

\begin{figure*}[htb]
\begin{center}
\subfigure[$n=1$]{\label{fig_L1}
\includegraphics[width=7cm]{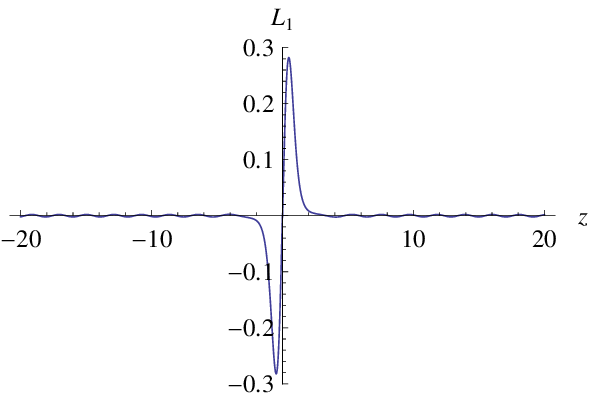}}
\subfigure[$n=2$]{\label{fig_L2}
\includegraphics[width=7cm]{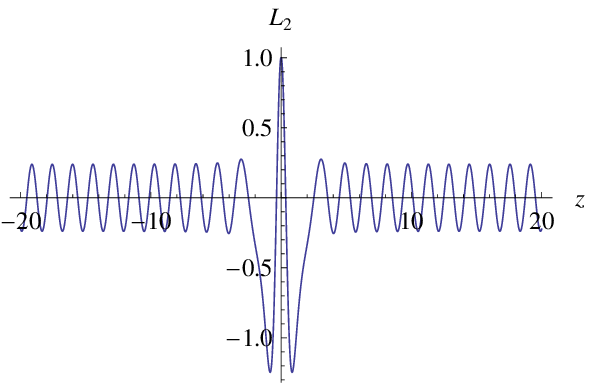}}
\end{center}\vskip -5mm
\caption{  The profiles of the wave function for the left-chiral
fermion resonances $L_{n}$ with the parameters $k=1$, $\xi=0.9$, $\eta=5$,
$\phi_{0}=1$ and $b=1$.}
 \label{fig_Ln}
\end{figure*}

\begin{figure*}[htb]
\begin{center}
\subfigure[$n=1$]{\label{fig_R1}
\includegraphics[width=7cm]{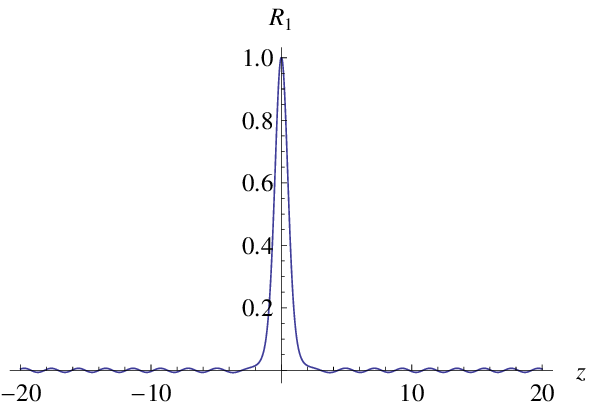}}
\subfigure[$n=2$]{\label{fig_R2}
\includegraphics[width=7cm]{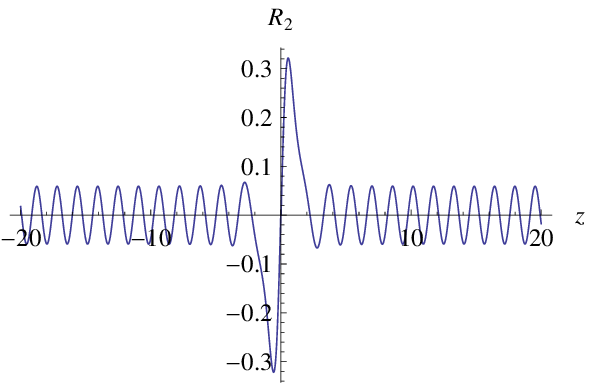}}
\end{center}\vskip -5mm
\caption{ The profiles of the wave function for the right-chiral
fermion resonances $R_{n}$ with the parameters $k=1$, $\xi=0.9$, $\eta=5$,
$\phi_{0}=1$ and $b=1$. }
 \label{fig_Rn}
\end{figure*}

For the single-kink background scalar field ($k=1$), the mass, width
and lifetime of the resonant KK modes with different values of $\xi$
are listed in Table \ref{tableFermink1}. As an example, we plot the
shapes of $P_{\text{L,R}}(m^{2})$ corresponding to $\xi=0.9$ and
$k=1$ in Fig.~\ref{fig_Pfk1c}. From Table~\ref{tableFermink1} and
Fig.~\ref{fig_Pfk1c}, it can be seen that the mass and lifetime of
left- and right-chiral fermion resonances are almost the same, so
the formation of the four-dimensional massive Dirac fermions can be
realized \cite{0901.3543}. The shapes of the resonances $L_{n}$ and
$R_{n}$ are shown in Figs.~\ref{fig_Ln} and \ref{fig_Rn} for
$\xi=0.9$ and $k=1$.
For the left-chiral KK resonances, it can be seen that the first
resonant KK mode is an odd-parity wave function, and the second one
has even parity. However, for the right-chiral KK resonances, the
first resonant KK mode has even parity and the second one has odd
parity.  This is held for any $n$-th fermion resonances, namely, the
parities of the $n$-th left- and right-chiral resonances are
opposite. In fact, this conclusion is originated from relationship
between the two potentials $V_L(z)$ and $V_R(z)$, or equivalently,
from the coupled equations of the left- and right-chiral fermions,
which are not given in this paper but one can refer
\cite{liu_0909.2312}. Further, four-dimensional massive Dirac
fermions can be obtained, which are consisted of the pairs of
coupled left- and right-chiral KK modes with different parities.

From Table \ref{tableFermink1}, it is shown that the lifetimes of
the first resonant KK modes satisfy
$\tau_{n=1}(\xi=0.5)>\tau_{n=1}(\xi=0.1)>\tau_{n=1}(\xi=0.9)$. From
Fig. \ref{fig_VfermionI}, it is seen that the width of the barrier
of the potentials in the vicinity of $V_{L,R}=m_{1}^{2}$ for
$\xi=0.5$ is close to that for $\xi=0.1$ and larger than that for
$\xi=0.9$, so $\tau_{n=1}(\xi=0.5)$ and $\tau_{n=1}(\xi=0.1)$ are
larger than $\tau_{n=1}(\xi=0.9)$. Moreover, the widths of the
barrier of the potentials for $\xi=0.5$ and $\xi=0.1$ are the same,
but the height of the barrier for $\xi=0.5$ is larger than that for
$\xi=0.1$, so $\tau_{n=1}(\xi=0.5)>\tau_{n=1}(\xi=0.1)$. For the
second resonant KK modes, the situation is similar to the first
ones. As Fig. \ref{fig_VfermionI} shown, the width of the barrier of
the potentials in the vicinity of $V_{L,R}=m_{2}^{2}$ for $\xi=0.5$
is the largest, so $\tau_{n=2}(\xi=0.5)$ is the largest one. Hence,
we can come to the conclusion that the lifetimes of the resonant KK
modes are decided by the width and the hight of the barrier of the
potential and the width is more important than the hight.

\begin{table*}[h]
\begin{center}
\renewcommand\arraystretch{1.3}
\begin{tabular}
 {|l|c|c|c|c|c|c|c|c|}
  \hline
 ~~~$\xi$~ & ~Chiral~ & ~Height of $V_{\text{L,R}}$~ & ~$n$~ & ~$m^{2}$~ & ~$m$~ & ~$\Gamma$~ & ~$\tau$~   \\
    \hline\hline
 $0.1$  & ~Left & $V_{L}^{\text{max}}=14.036$
     & ~$1$~   &~8.5206 ~& ~2.9190~ & ~0.001222~ & ~818.64177~ \\
    \cline{4-8}
   &  &  &
     ~$2$~ & ~13.6638~&~3.6965~  &~0.111288~  & ~8.98573~       \\
     \cline{2-8}
   & ~Right & $V_{R}^{\text{max}}=14.072$
     &~$1$~ &~8.5206~  & ~2.9190~  & ~0.001216~ & ~822.65478~\\
    \cline{4-8}
   &  &  &
     ~$2$~ & ~13.6463~& ~3.6941~  &~0.111813~  & ~8.94354~       \\
   \hline\hline
 $0.5$  & ~Left & $V_{L}^{\text{max}}=16.611$
     & ~$1$~   &~8.8961~& ~2.9826~ & ~0.000725~ & ~1380.15215~ \\
    \cline{4-8}
   &  &  &
     ~$2$~   & ~15.1313~&~3.8899~  &~0.063637~  & ~15.71416~       \\
     \cline{2-8}
   & ~Right & $V_{R}^{\text{max}}=16.651$
     &~$1$~ &~8.8961~  & ~2.9826~  & ~0.000722~ & ~1385.67372~\\
    \cline{4-8}
   &  &  &
     ~$2$~ & ~15.1188~& ~3.8883~  &~0.064661~  & ~15.46516~       \\
   \hline\hline
  $0.9$  & ~Left & $V_{L}^{\text{max}}=18.411$
     & ~$1$~   &~9.2353~& ~3.0390~ & ~0.003074~ & ~325.2705~ \\
    \cline{4-8}
   &  &  &
     ~$2$~   & ~16.3330~ & ~4.0414~  &~0.081694~  & ~12.2408~       \\
     \cline{2-8}
   & ~Right & $V_{R}^{\text{max}}=18.466$
     &~$1$~ &~9.2353~  & ~3.0390~  & ~0.003120~ & ~320.4962~\\
    \cline{4-8}
   &  &  &
     ~$2$~ & ~16.3363~& ~4.0418~  &~0.084267~  & ~11.8671~       \\
   \hline
\end{tabular}
\end{center}
\caption{The mass, width, and lifetime of resonances for fermions.
The parameters are set to $k=1$, $\eta=5$, $\phi_{0}=1$ and $b=1$.
Here $n$ is the order of resonant states with corresponding $m^2$
from small to large.} \label{tableFermink1}
\end{table*}

For double-kink background scalar field ($k=3$), because the mass,
width and lifetime of the right-chiral fermion resonant KK modes is
the same as that of left-chiral fermion resonances, we only list the
mass, width and lifetime for the left-chiral fermion resonant KK
modes with different values of $\xi$ are listed in Table
\ref{tableFermink3}. It is known that the number of the resonant KK
modes will increase when the coupling constant $\xi$ becomes larger.
It is also shown that the lifetimes of the first resonant KK modes
satisfy
$\tau_{n=1}(\xi=0.1)>\tau_{n=1}(\xi=0.3)>\tau_{n=1}(\xi=0.9)$, which
is because that the widths of the barrier of the potentials in the
vicinity of $V_{L,R}=m_{1}^{2}$ for $\xi=0.1$ are the largest and
that for $\xi=0.9$ are the smallest from Fig. \ref{fig_VfermionII}.
For the other resonant KK modes, the situation is similar to the
case of $k=1$.

\begin{table*}[h]
\begin{center}
\renewcommand\arraystretch{1.3}
\begin{tabular}
 {|c|c|c|c|c|c|c|c|c|}
  \hline
 $\xi$ & Chiral & ~Height of $V_{\text{L}}$~ & $n$ & ~$m^{2}$~ & ~$m$~ & ~$\Gamma$~ & ~$\tau$~   \\
    \hline\hline
 $0.1$  & ~Left & $V_{L}^{\text{max}}=15.648$
     & ~$1$~   &~3.1973 ~& ~1.7881~ & ~$2.92\times10^{-10}$~ & ~$3.42\times10^{9}$~ \\
    \cline{4-8}
       &    &
     & ~$2$~   &~8.8214~ & ~2.9701~ & ~$1.45\times10^{-4}$~ &~ $6.89\times10^{3}$~  \\
    \cline{4-8}
       &    &
     & ~$3$~   &~13.3115~ & ~3.6485~ & ~0.018426~ &~ 54.2705~  \\
   \hline\hline
 $0.3$  & ~Left & $V_{L}^{\text{max}}=17.160$
     & ~$1$~   &~3.2171 ~& ~1.79363~ & ~$3.91\times10^{-10}$~ & ~$2.56\times10^{9}$~ \\
    \cline{4-8}
       &    &
     & ~$2$~   &~8.9598~ & ~2.9933~ & ~$8.18\times10^{-5}$~ &~ $1.22\times10^{4}$~  \\
    \cline{4-8}
       &    &
     & ~$3$~   &~13.7770~ & ~3.7117~ & ~0.010634~ &~ 94.04994~  \\
   \hline\hline
 $0.9$  & ~Left & $V_{L}^{\text{max}}=20.395$
     & ~$1$~   &~3.2735 ~& ~1.8093~ & ~$9.03\times10^{-6}$~ & ~$1.11\times10^{5}$~ \\
    \cline{4-8}
       &    &
     & ~$2$~   &~9.3293~ & ~3.05439~ & ~$0.00114$~ &~ $878.85943$~  \\
    \cline{4-8}
       &    &
     & ~$3$~   &~14.9016~ & ~3.8603~ & ~0.01752~ &~ 57.07182~  \\
     \cline{4-8}
       &    &
     & ~$4$~   &~19.8003~ & ~4.44975~ & ~0.128982~ &~ 7.75318~  \\
   \hline
\end{tabular}
\end{center}
\caption{The mass, width, and lifetime of resonances for left-chiral
fermion. The parameters are set to $k=3$, $\eta=5$, $\phi_{0}=1$ and
$b=1$. Here $n$ is the order of resonant states with corresponding
$m^2$ from small to large.} \label{tableFermink3}
\end{table*}

\section{Conclusions and discussions}
\label{SecConclusion}

In this paper, we first review the model of thick
branes with a nonminimally coupled background scalar field and then
investigate the structure of the branes. In our model, the
background scalar field is set as single-kink ($k=1$) and
double-kink ($k=3$) solutions respectively. The nonminimally
coupling between the gravity and the background scalar field is
introduced via a term $\frac{1}{2}\xi R \phi^{2}$, and we find that
the behaviors of the warp factor and the branes are related to the
nonminimal coupling constant $\xi$. When the nonminimal coupling
constant $\xi$ is smaller than its critical value $\xi_0$, the
maximum of the warp factor is at $z=0$, and when $\xi>\xi_0$, the
maxima of the warp factor are at both sides of $z=0$. For the
single-kink ($k=1$) case, the brane will split into two sub-branes
with the increase of the nonminimal coupling constant $\xi$, and
the distance of the two sub-branes increases with $\xi$. For the
double-kink ($k=3$) case, there are two sub-branes, and their
distance also increases with $\xi$. The scalar potential $V(\phi)$
is a double well potential, and two vacua of the potential are at
$\pm\phi_0$ for $\xi=0$ and not at $\pm\phi_0$ for $\xi>0$,
respectively. Moreover, we investigate the effects of the
nonminimal coupling constant on the localization of gravity and
various bulk matter fields on the branes.

Firstly, the localization of gravity is considered. It is found
that, for the case of single-kink ($k=1$), the gravity zero mode is
localized on the single brane with small $\xi$, and it is localized
on the center of the two sub-branes with large $\xi$. For the case
of double-kink ($k=3$), the gravity zero mode is localized between
the two sub-branes. All the massive modes are continuous spectrum
wave functions and could not be localized on the branes. However,
for larger coupling constant $\xi$, there could exist gravity
resonant states, and the number of the resonances increases with
$\xi$.

For scalar field, the zero mode is localized on the center of the
branes ($k=1$ case) or between them ($k=3$ case) when $\xi<\xi_{0}$,
while it is localized on each sub-brane when $\xi>\xi_{0}$. For the
vector field, the zero mode can not be localized on the branes. All
the massive modes for scalars and vectors are continuous spectrum
wave functions and can not localized on the branes. There is not
exist the resonant state for both fields.

For spin-1/2 fermion fields, in order to localized the left- and
right-chiral fermions, we introduce the usual Yukawa coupling
$\eta\bar{\Psi}\phi\Psi$. We find that for positive Yukawa coupling
constant $\eta$ larger than its critical value $\eta_{0}$, the
left-chiral fermion zero mode can be localized on the branes. For
the case of $k=1$, the left-chiral fermion zero is localized on the
single brane with small $\xi$ and on the center of the two
sub-branes with large $\xi$. For the case of $k=3$, the left-chiral
fermion zero is localized between the two sub-branes. The massive KK
modes asymptotically turn into continuous plane waves when far away
from the branes. It is interesting that the well of the potentials
for left- and right-chiral fermions becomes deeper and deeper with
the increase of the nonminimal coupling constant $\xi$, which leads
to a series of massive fermions with a finite lifetime on the
branes. The spectra of left- and right-chiral fermion resonances are
the same, which demonstrates that a Dirac fermion with a finite
lifetime on the branes can be composed from the left- and
right-chiral fermion resonant KK modes.

At the end of this paper, we discuss the effect of the
nonminimal coupling constant $\xi$ on the localization of gravity
and various matter fields. For these zero modes, the situation is
similar to the case of minimal coupling, i.e., only the vector zero
mode is not localized on the branes, and all the other zero modes
are localized on the branes. This indicates that the nonminimal
coupling constant $\xi$ does not effect the localization of these
zero modes. It is because that the localization of the zero modes is
decided by the behavior of the system at
the infinity of the extra dimension. The behavior of the system for
the case of the nonminimal coupling, is similar to the one for the
case of minimal coupling when far away from the branes. This can
also be seen from the action (\ref{action}). When the extra dimensional
coordinate tends to infinity, $\phi(\pm\infty)\rightarrow\pm\phi_0$, and the action
(\ref{action}) is reduced to the following form
\begin{eqnarray}
 S = \int d^5 x \sqrt{-g}\left [ \frac{\zeta}{2\kappa_{5}^{2}}
     R -\frac{1}{2}
     g^{MN}\partial_M \phi \partial_N \phi - V(\phi) \right ]
\label{action_infinity}
\end{eqnarray}
with the constant $\zeta=(1-\kappa_{5}^{2}\xi\phi_{0}^{2})$, which
is similar to the general action of the thick brane for the case of
minimal coupling. Therefore, the behavior of the zero modes is
similar to these for the case of minimal coupling at
$z\rightarrow\pm\infty$. However, near the branes, the effects of the nonminimal coupling constant
$\xi$ are very obvious for the localization of gravity and matter
fields. For the scalar zero mode, when $\xi$ is large, the zero mode
is localized on each sub-brane. For the gravity and the fermion
field, the number of the resonances increases with $\xi$.

\section*{Acknowledgement}

The authors are extremely grateful for the anonymous referee, whose
comments led to the improvement of this paper. This work was
supported by the Program for New Century Excellent Talents in
University, the Huo Ying-Dong Education Foundation of Chinese
Ministry of Education (Grant No. 121106), the National Natural Science
Foundation of China (Grant No. 11075065), the Doctoral Program
Foundation of Institutions of Higher Education of China (Grant No.
20090211110028), and the Fundamental Research Funds for the Central Universities (Grant No. lzujbky-2012-k30).

\end{document}